\documentclass{article}

\usepackage{arxiv}

\usepackage[utf8]{inputenc} 
\usepackage[T1]{fontenc}    
\usepackage{hyperref}       
\usepackage{url}            
\usepackage{booktabs}       
\usepackage{amsfonts}       
\usepackage{nicefrac}       
\usepackage{microtype}      
\usepackage{lipsum}

\usepackage{graphicx}
\usepackage{xcolor}
\usepackage{amsmath}
\usepackage{tikz}
\usepackage{pgfplots}
\usepackage{subfig}
\usepackage{comment}
\usepackage{appendix}

\title{Adaptive Mesh Refinement and Coarsening for Diffusion-Reaction Epidemiological Models}

\author{
  Malú Grave \\
  Dept. of Civil Engineering\\
  COPPE/Federal University of Rio de Janeiro \\
  P.O. Box 68506, RJ 21945-970, Rio de Janeiro, Brazil \\
  \texttt{malugrave@nacad.ufrj.br} \\
   \And
 Alvaro L.G.A. Coutinho \\
  Dept. of Civil Engineering\\
  COPPE/Federal University of Rio de Janeiro \\
  P.O. Box 68506, RJ 21945-970, Rio de Janeiro, Brazil \\
  \texttt{alvaro@nacad.ufrj.br} \\
}

\begin{document}
\maketitle

\begin{abstract}
The outbreak of COVID-19 in 2020 has led to a surge in the interest in the mathematical modeling of infectious diseases. Disease transmission may be modeled as compartmental models, in which the population under study is divided into compartments and has assumptions about the nature and time rate of transfer from one compartment to another. Usually, they are composed of a system of ordinary differential equations (ODE's) in time. A class of such models considers the Susceptible, Exposed, Infected, Recovered, and  Deceased populations, the SEIRD model. However, these models do not always account for the movement of individuals from one region to another. In this work, we extend the formulation of SEIRD compartmental models to diffusion-reaction systems of partial differential equations to capture the continuous spatio-temporal dynamics of COVID-19. Since the virus spread is not only through diffusion, we introduce a source term to the equation system, representing exposed people who return from travel. We also add the possibility of anisotropic non-homogeneous diffusion. We implement the whole model in \texttt{libMesh}, an open finite element library that provides a framework for multiphysics, considering adaptive mesh refinement and coarsening.  Therefore, the model can represent several spatial scales, adapting the resolution to the disease dynamics. We verify our model with standard SEIRD models and show several examples highlighting the present model's new capabilities.
\end{abstract}


\section{Introduction}

The COVID-19 pandemic has caused widespread damage worldwide, in terms of human lives and international economic weakening. As a new highly contagious disease, governments have taken unprecedented measures to slow the spread of the virus, including quarantines, curfews, lockdowns, and national and international travel suspension. These measures, considered essential by many experts, are partly motivated by the lack of reliable data on this disease's transmission and lethality, which justifies cautious responses from the authorities and population. These events demonstrate more than ever the need for reliable tools designed to model the spatio-temporal spread of infectious diseases.

The study of infectious disease proliferation is a well-established field and has given rise to the area of science called \textit{mathematical epidemiology} \cite{yang2001epidemiologia}. Mathematical epidemiology proposes models that help the understanding of epidemics and to outline policies to control infectious diseases. In Brazil, studies of this type have been carried out for years for diseases such as Dengue \cite{erickson2010dengue} and Zika \cite{dantas2018calibration}, and, in a global context, diseases such as HIV \cite{mukandavire2010global}, SARS \cite{dye2003modeling}, Malaria \cite {midekisa2012remote}, Ebola \cite {lekone2006statistical}, among others. The COVID-19 pandemic brought the need for more research in this area. Several models for this pandemic outbreak have been presented in recent months \cite{viguerie2020diffusion, viguerie2020simulating, arino2020simple, giordano2020modelling, carcione2020simulation,volpatto2020spreading}.

Disease transmission may be modeled as \textit{compartmental models}, in which the population under study is divided into compartments and has assumptions about the nature and time rate of transfer from one compartment to another \cite{brauer2019mathematical}. These models have been used extensively in biological, ecological, and chemical applications \cite{brauer2012mathematical, keller2013numerical, jourdan2019compartmental}. They allow for an understanding of the processes at work and for predicting the dynamics of the epidemic.

The large majority of the compartmental models are composed of a system of ordinary differential equations (ODE's) in time. Though compartmentalized models are simple to formulate, analyze,
and solve numerically, these models do not always account for the movement of individuals from one region to another. Different approaches have been used to introduce spatial variation into such ODE models \cite{holmes1994partial, keeling2011modeling, gatto2020spread, giordano2020modelling}. The strategies consist of defining regional compartments corresponding to different geographic units and adding coupling terms to the model equations to account for species' movement from unit to unit.

In this work, we use a partial differential equation (PDE) model to capture the continuous spatio-temporal dynamics of COVID-19. PDE models incorporate spatial information more naturally and allow for capture the dynamics across several scales of interest. They have a significant advantage over ODE models, whose ability to describe spatial information is limited by the number of geographic compartments. Indeed, recent research indicates that COVID-19 spreading presents multi-scale features that go from the virus and individual immune system scale to the collective behavior of a whole population \cite{bellomo2020multi}. We study a compartmental SEIRD model \textit{(susceptible, exposed, infected, recovered, deceased)} that incorporates spatial spread through diffusion terms \cite{keller2013numerical, kim1996galerkin, viguerie2020diffusion, viguerie2020simulating, cantrell2004spatial}. Adaptive mesh refinement and coarsening \cite{carey1997} can resolve population dynamics from local (street, city) to regional (district, state), providing an accurate spatio-temporal description of the infection spreading. Moreover, diffusion may be properly tuned to account for local natural or social inhomogeneities (e.g., mountains, lakes, highways) describing populations' movements.

However, the main limitation of the diffusion-reaction PDE approach is the definition of the diffusion operator and transmission coefficients, which depend on the population's behavior. Another issue is that the virus spread is not only through diffusion, since people, who may be infected, travel long distances in a short period. Some models relate the mobile geolocation data with the spread of the disease \cite{peixoto2020modeling, kraemer2020effect}.  These issues make the model a highly complex system, which may completely change as the population's behavior changes.  Therefore, this work contributes to improving the knowledge of compartmental diffusion-reaction PDE models.

All implementations are done using the \texttt{libMesh} library. As other freely available open-source libraries (deal.II \cite{bangerth2007deal}, FEniCS \cite{AlnaesBlechta2015a}, GRINS \cite{bauman2016grins}, MOOSE \cite{gaston2009moose}, etc), \texttt{libmesh} provides a finite element framework that can be used for numerical simulation of partial differential equations in various applied fields in science and engineering. It has already been used in more than 1,000 publications with applications in many different areas. See, for example, recent applications in sediment transport \cite{grave2020residual} and bubble dynamics \cite{Grave_Camata_Coutinho_2020}. This library is an excellent tool for programming the finite element method and can be used for one-, two-, and three-dimensional steady and transient simulations on serial and parallel platforms. The \texttt{libmesh} library provides native support for adaptive mesh refinement and coarsening, thus providing a natural environment for the present study. The main advantage of this library is the possibility of focusing on implementing the specifics features of the modeling without worrying about adaptivity and code parallelization. Consequently, the effort to build a high performance computing code tends to be minimized. about adaptivity and code parallelization.

The remainder of this work is organized as follows: In section \ref{GE}, we present the governing equations that describe the dynamics of a virus infection. First, we present a generic spatio-temporal SEIRD model, based on the EPIDEMIC software \cite{EPIDEMICmanual}, used to verify our implementation. We then present a model that better represents the dynamics of COVID-19 infection spread, based on \cite{viguerie2020simulating, viguerie2020diffusion}. In section \ref{FEF}, we introduce the Galerkin finite element formulation, the time discretization, and the \texttt{libMesh} implementation. Then, we present the numerical verification of the generic spatio-temporal SEIRD model implementation. We verify our algorithm's capacity to represent a compartmental model \cite{EPIDEMICmanual} and show how the diffusion influences the dynamics. Section \ref{NRC19} presents the numerical results of the spatio-temporal model of COVID-19 infection spread. We perform simulations similar to the ones presented in \cite{viguerie2020diffusion} and show tests to highlight the new modeling capabilities introduced in this work. Finally, the paper ends with a summary of our main findings and the perspectives for the next steps of this research.

\section{Governing equations}\label{GE}

The presentation of the governing equations follows the  continuum mechanics framework in \cite{viguerie2020diffusion} instead of the more traditional approach found in mathematical and biological references. Consider a system which may be decomposed into $N$ distinct populations: $u_1(\mathbf{x}, t)$, $u_2(\mathbf{x}, t)$, ..., $u_N (\mathbf{x}, t)$.  Let $\Omega \in R^2$ be a simply connected domain of interest with boundary $\partial \Omega = \Gamma_D \cap \Gamma_N$, and $[0, T]$ a generic time interval. The vector compact representation of the governing equations as a transient nonlinear diffusion-reaction system of equations reads,

\begin{equation}\label{eq:system}
    \frac{\partial\mathbf{u}}{\partial t} + \left(\mathbf{A} + \mathbf{B(u)}\right)\mathbf{u} - \nabla \cdot (\boldsymbol{\nu} \nabla\mathbf{u}) - \mathbf{f}=0 \textrm{ in }\Omega \times [0,T] 
\end{equation}

\begin{equation}\label{eq:direchlet}
\mathbf{u}=\mathbf{u_D} \textrm{ in }\Gamma_D \times [0,T] 
\end{equation}

\begin{equation}\label{eq:neumann}
(\boldsymbol{\nu} \nabla\mathbf{u}) \cdot \mathbf{n}=\mathbf{h} \textrm{ in }\Gamma_N \times [0,T] 
\end{equation}

We denote the densities of the \textit{susceptible}, \textit{exposed}, \textit{infected}, \textit{recovered} and \textit{deceased} populations as $s(\mathbf{x}, t)$, $e(\mathbf{x}, t)$, $i(\mathbf{x}, t)$, $r(\mathbf{x}, t)$, and $d(\mathbf{x}, t)$. Also, let $c(\mathbf{x}, t)$ denote the \textit{cumulative number of infected} and 
$n(\mathbf{x}, t)$ the sum of the living population; i.e.,
$n(\mathbf{x}, t) = s(\mathbf{x}, t) + e(\mathbf{x}, t) + i(\mathbf{x}, t) + r(\mathbf{x}, t)$. We consider $\mathbf{u} = [s,e,i,r,d]^T$. The matrices $\mathbf{A}$, $\mathbf{B}$ and $\boldsymbol{\nu}$, and the vector $\mathbf{f}$ depend on a particular form of the system dynamics. Furthermore, in general, $\boldsymbol{\nu}=\boldsymbol{\nu}(\mathbf{x})$, that is, diffusion is heterogeneous and anisotropic. Besides the boundary  conditions \eqref{eq:direchlet}, \eqref{eq:neumann}, we specify the initial condition $\mathbf{u}(\mathbf{x},0)=\mathbf{u}_0$. The total population $U_i(t)$ of each compartment $u_i(\mathbf{x}, t)$ is,

\begin{equation}
    U_i(t) = \int_\Omega u_i(\mathbf{x}, t) d\Omega
\end{equation}

\noindent for $i=1,2, \cdots, N$.

\subsection{Generic spatio-temporal SEIRD model}

We first consider a SEIRD model \cite{brauer2019mathematical} given by the following system of coupled PDEs over $\Omega \times [0, T]$:

\begin{equation}
\frac{\partial s}{\partial t} + \frac{\beta}{n} si - \nabla \cdot (n\nu_s\nabla s) = 0
\label{epidemic_s}
\end{equation}
\begin{equation}
\frac{\partial e}{\partial t} - \frac{\beta}{n} si + \alpha e - \nabla \cdot (n\nu_e \nabla e) = 0
\label{epidemic_e}
\end{equation}
\begin{equation}
\frac{\partial i}{\partial t} - \alpha e + (\gamma  + \delta) i - \nabla \cdot (n\nu_i \nabla i) = 0
\label{epidemic_i}
\end{equation}
\begin{equation}
\frac{\partial r}{\partial t}  - \gamma i  -\nabla \cdot (n \nu_r \nabla r) = 0
\label{epidemic_r}
\end{equation}
\begin{equation}
\frac{\partial d}{\partial t} - \delta i = 0
\label{epidemic_d}
\end{equation}

\noindent where $\beta$ is transmission rate $(days^{-1})$, $\alpha$ the latent rate $(days^{-1})$, $\gamma$ the recovery rate $(days^{-1})$, $\delta$ the death rate $(days^{-1})$, and $\nu_s$, $\nu_e$, $\nu_i$, $\nu_r$ are diffusion parameters respectively corresponding to the different population groups ($km^2  persons^{-1}  days^{-1}$). We append to the system of equations homogeneous Neumann boundary conditions, that is, $(\boldsymbol{\nu} \cdot \nabla\mathbf{u}) \cdot \mathbf{n} = 0$.

We can reframe this model in the general form given by equation \eqref{eq:system}. Thus, the matrices $\mathbf{A}$, $\mathbf{B}$, $\boldsymbol{\nu}$ and the vector $\mathbf{f}$ reads,

\begin{equation}
   \mathbf{A} = \begin{bmatrix}
0 & 0 & 0 & 0 & 0\\
0 & \alpha & 0 & 0 & 0\\
0 & -\alpha & \gamma+\delta & 0 & 0\\
0 & 0 & -\gamma & 0 & 0\\
0 & 0 & -\delta & 0 & 0\\
\end{bmatrix}
\end{equation}
\begin{equation}
   \mathbf{B} = \begin{bmatrix}
0 & 0 & \frac{\beta}{n}s &  0  & 0\\
0 & 0 & -\frac{\beta}{n}s & 0 & 0\\
0 & 0 & 0 & 0 & 0\\
0 & 0 & 0 & 0 & 0\\
0 & 0 & 0 & 0 & 0\\
\end{bmatrix}
\end{equation}
\begin{equation}
   \boldsymbol{\nu} = \begin{bmatrix}
\boldsymbol{\nu_s} & 0 & 0 & 0  & 0\\
0 & \boldsymbol{\nu_e} & 0 & 0 & 0\\
0 & 0 & \boldsymbol{\nu_i} & 0 & 0\\
0 & 0 & 0 & \boldsymbol{\nu_r} & 0\\
0 & 0 & 0 & 0 & 0\\
\end{bmatrix}
\end{equation}
\begin{equation}
   \boldsymbol{\nu^k} = \begin{bmatrix}
 \nu^k_{xx} & \nu^k_{xy}\\
 \nu^k_{yx} & \nu^k_{yy} 
\end{bmatrix}
\textnormal{ with } k=s,e,i,r
\end{equation}
\begin{equation}
   \mathbf{f} = \mathbf{0}
\end{equation}

This model is based on the EPIDEMIC software\footnote{https://americocunhajr.github.io/EPIDEMIC/ \cite{EPIDEMICmanual}}, and it is employed to verify our implementation. The system of equations represents that the susceptible population decreases as the exposed population increases. This variation depends on the transmission rate between infected and susceptible. The number of exposed increases because of the transmission rate and decreases when the exposed individuals become infected (after the incubation period). The number of infected increases after the incubation period and decreases depending on the recovery and death rate. The number of deaths depends only on the death rate as the number of recovered depends only on the recovery rate. Finally, the cumulative number of infected depends only on the exposed and the incubation period. The diffusion parameters are included in the model to spread the disease spatially.

Summarizing, this model assumes:

\begin{itemize}
    \item Movement is proportional to population size; i.e., more movement occurs within heavily populated regions;
    \item No movement occurs among the deceased population;
    \item There is a latency period between exposure and the development of symptoms;
    \item The probability of contagion is inversely proportional to the population size;
    \item The exposed persons will ever develop symptoms;
    \item Only infected persons are capable of spreading the disease;
    \item The non-virus mortality rate is not considered in this model;
    \item New births are not considered in this model.
\end{itemize}

Note that the EPIDEMIC model's dynamics does not represent the actual COVID19 dynamics since, in the case of COVID19, the exposed population may be asymptomatic and recover without becoming infected and still spread the virus. Thus, a better model would be the one based on \cite{viguerie2020simulating, viguerie2020diffusion}.

\subsection{Spatio-temporal model of COVID-19
infection spread}

We begin by making several model assumptions to represent the COVID-19 infection spread adequately \cite{viguerie2020diffusion}:

\begin{itemize}
    \item Only mortality due the COVID-19 is considered;
    \item New births are not considered in this model.
    \item Some portion of exposed persons never develop symptoms, and move directly from the exposed compartment to the recovered compartment (asymptomatic cases);
    \item Both asymptomatic (exposed) and symptomatic (infected) patients are capable of spreading the disease;
    \item There is a latency period between exposure and the development of symptoms;
    \item It is possible that new cases of exposed people appear randomly in the system (exposed people who return from a travel);
    \item The probability of contagion increases with population size (Allee effect \cite{viguerie2020simulating});
    \item Movement is proportional to population size; i.e., more movement occurs within heavily populated regions;
    \item No movement occurs among the deceased population;

\end{itemize}

Then, the system of equations becomes:

\begin{equation}
\frac{\partial s}{\partial t} + \beta_i \left(1-\frac{A}{n}\right)si + \beta_e \left(1-\frac{A}{n}\right)se +f - \nabla \cdot (n\nu_s\nabla s) = 0
\label{covid_s}
\end{equation}
\begin{equation}
\frac{\partial e}{\partial t} - \beta_i \left(1-\frac{A}{n}\right)si - \beta_e \left(1-\frac{A}{n}\right)se + (\alpha  + \gamma_e) e -f- \nabla \cdot (n\nu_e \nabla e) = 0
\label{covid_e}
\end{equation}
\begin{equation}
\frac{\partial i}{\partial t} - \alpha e + (\gamma_i  + \delta) i - \nabla \cdot (n\nu_i \nabla i) = 0
\label{covid_i}
\end{equation}
\begin{equation}
\frac{\partial r}{\partial t}  -\gamma_e e- \gamma_i i  -\nabla \cdot (n \nu_r \nabla r) = 0
\label{covid_r}
\end{equation}
\begin{equation}
\frac{\partial d}{\partial t} - \delta i = 0
\label{covid_d}
\end{equation}

\noindent where $A$ characterizes the Allee effect ($persons$), that takes into account the tendency of outbreaks to cluster around large populations, $\beta_i$ is the transmission rate between symptomatic and susceptible $(persons^{-1}days^{-1})$, $\beta_e$ is the transmission rate between asymptomatic and susceptible $(persons^{-1}days^{-1})$, $f$ is a source function that depends on space and time $(persons)$, $\alpha$ is the latent rate $(days^{-1})$, $\gamma_e$ is the recovery rate of the asymptomatic $(days^{-1})$, $\gamma_i$ is the recovery rate of the symptomatic $(days^{-1})$, $\delta$ is the death rate $(days^{-1})$, and $\nu_s$, $\nu_e$, $\nu_i$, $\nu_r$ are the diffusion parameters respectively corresponding to the different population groups ($km^2  persons^{-1}  days^{-1}$).

Now, we call \textit{exposed} who has contact with the virus but remains asymptomatic. However, since the virus is highly transmissible, the exposed population also may transmit the virus. The exposed may recover without any symptoms or may become \textit{infected}. The infected follow the same logic of the previous SEIRD system (they may recover or die). The main difference in the new SEIRD system is in the exposed population and whom it interacts. The source function $f$ may be defined to represent exposed people who return from travel. Note that $\beta$ has units ($days^{-1}$) while $\beta_i$ and $\beta_e$ have units ($person^{-1}days^{-1}$). 
While equations (\ref{epidemic_s}) and (\ref{epidemic_e}) divide $\beta$ by the living population, equations (\ref{covid_s}), (\ref{covid_e}) and (\ref{covid_i}) keep $\beta_i$ and $\beta_e$ constant independent of that.

Therefore, to express this model in the general form given by equation \eqref{eq:system}, the matrices $\mathbf{A}$, $\mathbf{B}$, $\boldsymbol{\nu}$ and the vector $\mathbf{f}$ reads,

\begin{equation}
   \mathbf{A} = \begin{bmatrix}
0 & 0 & 0 & 0 & 0\\
0 & \alpha+\gamma_e & 0 & 0 & 0\\
0 & -\alpha & \gamma_i+\delta & 0 & 0\\
0 & -\gamma_e & -\gamma_i & 0 & 0\\
0 & 0 & -\delta & 0 & 0\\
\end{bmatrix}
\end{equation}
\begin{equation}
   \mathbf{B} = \begin{bmatrix}
0 & \beta_e\left(1-\frac{A}{n}\right)s & \beta_i\left(1-\frac{A}{n}\right)s & 0  & 0\\
0 & -\beta_e\left(1-\frac{A}{n}\right)s & -\beta_i\left(1-\frac{A}{n}\right)s & 0 & 0\\
0 & 0 & 0 & 0 & 0\\
0 & 0 & 0 & 0 & 0\\
0 & 0 & 0 & 0 & 0\\
\end{bmatrix}
\end{equation}
\begin{equation}
   \boldsymbol{\nu} = \begin{bmatrix}
\boldsymbol{\nu_s} & 0 & 0 & 0  & 0\\
0 & \boldsymbol{\nu_e} & 0 & 0 & 0\\
0 & 0 & \boldsymbol{\nu_i} & 0 & 0\\
0 & 0 & 0 & \boldsymbol{\nu_r} & 0\\
0 & 0 & 0 & 0 & 0\\
\end{bmatrix}
\end{equation}
\begin{equation}
   \boldsymbol{\nu^k} = \begin{bmatrix}
 \nu^k_{xx} & \nu^k_{xy}\\
 \nu^k_{yx} & \nu^k_{yy} 
\end{bmatrix}
\textnormal{ with } k=s,e,i,r
\end{equation}
\begin{equation}
   \mathbf{f} = \begin{bmatrix}
-f\\
f \\
0 \\
0 \\
0 \\
\end{bmatrix}
\end{equation}


 If we assume that the region of interest is isolated, we prescribe the following homogeneous Neumann boundary conditions,
 
 \begin{equation}
     \nabla s \cdot \mathbf{n} = 0
 \end{equation}
 \begin{equation}
     \nabla e \cdot \mathbf{n} = 0
 \end{equation}
  \begin{equation}
     \nabla i \cdot \mathbf{n} = 0
 \end{equation}
  \begin{equation}
     \nabla r \cdot \mathbf{n} = 0
 \end{equation}
 
\noindent or simply $(\boldsymbol{\nu} \cdot \nabla\mathbf{u}) \cdot \mathbf{n} = 0$.

\subsection{Determination of $R_0$}

The basic reproduction number, $R_0$, is defined as the average number of additional infections produced by an infected individual in a wholly susceptible population over the full course of the disease outbreak. In an epidemic situation, the threshold $R_0 = 1$ is the dividing line between the infection dying out and the onset of an epidemic. $R_0 > 1$ implies growth of the epidemic, whereas $R_0 < 1$ implies decay in infectious spread \cite{brauer2019mathematical}.

The concept of $R_0$ is well-defined for ODE models. However, its extension to a PDE model is unclear, owing to the influence of diffusion. Viguerie et al. \cite{viguerie2020diffusion} found that a $R_0$ derived for the ODE version of the PDE model is not consistently reliable to represent the epidemic's dynamic growth adequately. If we do not consider the diffusion, $R_0$ may be calculated as:

\begin{equation}
    R_0 = \frac{\beta_es + f}{\alpha + \gamma_e} + \frac{\beta_i\alpha s}{(\alpha+\gamma_e)(\delta +\gamma_i)}
\end{equation}

For further details about the $R_0$ calculation, refer to \cite{diekmann1990definition, viguerie2020diffusion}.

\section{Finite Element Formulation}\label{FEF}

In this section we briefly introduce the Galerkin finite element formulation, the time discretization, and the the \texttt{libMesh} implementation, supporting adaptive mesh refinement and coarsening. Appendices A and B give respectively the resulting finite element matrices for the generic spatio-temporal SEIRD and COVID-19 models. 

\subsection{Space Discretization}

We introduce a Galerkin finite element variational formulation for space discretization.  Without loss of generality, we consider the case of homogeneous Dirichlet and Neumann boundary conditions. Let $\mathbf{V_u}^h$ be a finite dimensional space such that,

\begin{equation}
    \mathbf{V_u}^h=\{\mathbf{u}^h(\cdot,t), \mathbf{w}^h(\cdot) \in H^1(\Omega) \enspace |\enspace \mathbf{u}^h=0, \mathbf{w}^h=0 \textnormal{ on } \Gamma_D\}
\end{equation}

\noindent in which $\mathbf{u}^h(\cdot,t)$ is the discrete counterpart of $\mathbf{u}$ and $\mathbf{w}^h$ the weight function. The weak formulation is then: find $\mathbf{u}^h \in \mathbf{V_u}^h$ such that $\forall \mathbf{w}^h \in \mathbf{V_u}^h$,

\begin{equation}
    \left(\mathbf{w}^h,\frac{\partial\mathbf{u}^h}{\partial t}\right) + \bigg(\mathbf{w}^h, \left(\mathbf{A} + \mathbf{B(u}^h)\right)\mathbf{u}^h\bigg) - \bigg(\mathbf{w}^h,\nabla \cdot (\boldsymbol{\nu} \nabla\mathbf{u}^h)\bigg) - \bigg(\mathbf{w}^h,\mathbf{f}\bigg)=0 \;  \textrm{ in }\Omega \times [0,T]
\end{equation}

\begin{equation}
\left(\mathbf{w}^h, \mathbf{u}^h(\cdot,0) \right) = \left(\mathbf{w}^h, \mathbf{u}_0 \right)  \textrm{ in }\Omega
\end{equation}

Here we define the operation $(\cdot,\cdot)$ as the standard scalar product in $L^2(\Omega)$.

\subsection{Time Integration}

The SEIRD and COVID-19 models yield stiff systems of equations, making explicit time-marching methods unfeasible. The Backward Euler method is widely applied because of its unconditional numerical stability characteristics. However, it has the disadvantage of being only first-order accurate, which introduces a significant amount of numerical diffusion. Then, we use the second-order Backward Differentiation Formula (BDF2), which, compared to the prevailing Backward Euler method, has significantly better accuracy while retaining unconditional linear stability. The model becomes,

\begin{equation}
\begin{split}
    \left(\mathbf{w}^h,\frac{1.5\mathbf{u}_{n+1}^h-2\mathbf{u}^h_n +  0.5\mathbf{u}^h_{n-1}}{\Delta t}\right) + \bigg(\mathbf{w}^h, \left(\mathbf{A} + \mathbf{B(u}_{n+1}^h)\right)\mathbf{u}_{n+1}^h\bigg) \\
    - \bigg(\mathbf{w}^h,\nabla \cdot (\boldsymbol{\nu} \nabla\mathbf{u}_{n+1}^h)\bigg) - \bigg(\mathbf{w}^h,\mathbf{f}_{n+1}\bigg)=0 \textrm{ in }\Omega \times [0,T]
\end{split}
\end{equation}

The subscript ${n+1}$ is associated to $t=t_{n+1}$ and $n$, and ${n-1}$ to the previous time-steps. 

\subsection{Implementation and Adaptive Mesh Refinement}

We implement the compartmental epidemiological models  in \texttt{libMesh}, a C++ FEM open-source software library for parallel adaptive finite element applications \cite{libmesh}. \texttt{libMesh} also interfaces with external solver packages like PETSc \cite{petsc-web-page} and Trilinos \cite{trilinos-website}. Recently, \texttt{libMesh} was also coupled with in-situ visualization and data-analysis tools \cite{camata2018situ, silva2020dfanalyzer}. It provides a finite element framework that can be used for the numerical simulation of partial differential equations on serial and parallel platforms. This library is an excellent tool for programming the finite element method and can be used for one-, two-, and three-dimensional steady and transient simulations.  The \texttt{libMesh} library also has native support for adaptive mesh refinement and coarsening (AMR/C).

Multiple scales can be resolved by AMR/C. \texttt{libMesh} supports AMR/C by $h$-refinement (element subdivision), $p$-refinement (increasing the polynomial approximation order), and $hp$-refinement, that is, a combination of both \cite{carey1997}. In \texttt{libMesh}, coarsening is supported in the $h$, $p$, and $hp$ AMR/C options. In the present work, we restrict ourselves to $h$-refinement with hanging nodes. The AMR/C procedure uses a local error estimator to drive the refinement and coarsening procedure, considering the error of an element relative to its neighbor elements in the mesh. This error may come from any variable of the system. As it is standard in \texttt{libMesh}, Kelly$'$s error indicator is employed, which uses the $H^1$ seminorm to estimate the error \cite{ainsworth}. Apart from the element interior residual, the flux jumps across the inter-element edges influence the element error. The flux jump of each edge is computed and added to the element error contribution. For both the element residual and flux jump, the values of the desired variables at each node are necessary. Therefore, the error $\left\lVert e \right\rVert^2$ can be stated as,
\begin{equation}
\left\lVert e \right\rVert^2 = \sum_{i=1}^n \left\lVert e \right\rVert^2_i 
\end{equation}
\noindent where $\left\lVert e \right\rVert^2_i$ is the error of each variable. In this study, we use all population types as variables for the error estimator.

\begin{figure}[htpb]
    \centering
    \includegraphics[width=0.8\linewidth]{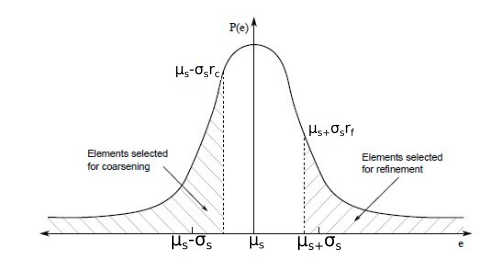}
    \caption{Statistical refinement strategy - elements in hatched areas are flagged to AMR/C process.}
    \label{fig:meshprobability}
\end{figure}

\begin{figure}[htpb]
    \centering
    \includegraphics[width=0.6\linewidth]{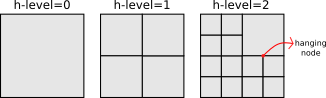}
    \caption{Adaptive mesh refinement: hierarchy of refined meshes with hanging nodes, where the solution is constrained to enforce continuity.}
    \label{fig:hlevel}
\end{figure}

After computing the error values, the elements are ‘‘flagged'' for refining and coarsening regarding their relative error. This is done by a statistical element flagging strategy. It is assumed
that the element error $\left\lVert e \right\rVert$ is distributed approximately in a normal probability function. Here, the statistical mean $\mu_s$ and standard deviation $\sigma_s$ of all errors are calculated. Whether an element
is flagged is depending on a refining ($r_f$) and a coarsening ($c_f$) fraction. For all errors $\left\lVert e \right\rVert < \mu_s - \sigma_s c_f$ the elements are flagged for coarsening and for all $\left\lVert e \right\rVert > \mu_s + \sigma_s r_f$ the elements are marked for refinement (see Figure \ref{fig:meshprobability}). 
The refinement level is limited by a maximum $h$-level
($h_{max}$), (see Figure \ref{fig:hlevel}), and the coarsening is done by $h$-restitution of sub-elements \cite{carey1997}, \cite{kelly}. 

\section{Numerical Results: Verification of the generic spatio-temporal SEIRD model}\label{NRGS}

To verify the implementation of the generic spatio-temporal SEIRD model, we have done several tests. For this, we consider a square domain of $1km\times1km$ centered at $(0,0)$ for all tests in this section. 

\subsection{Test 1: Reproducing  a compartmental model}

\begin{figure}[htpb]
    \centering
    \includegraphics[width=0.8\linewidth]{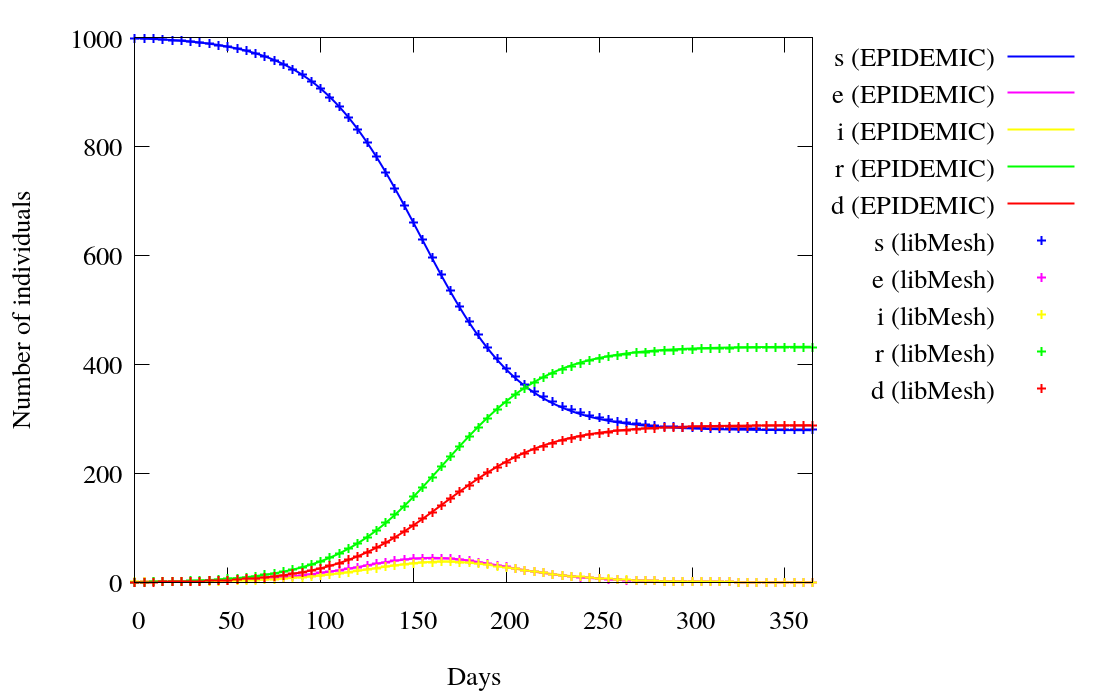}    
    \caption{Test 1: Reproducing a compartmental model}
    \label{fig:seird_epidemic}
\end{figure}

In the first test, we do not consider diffusion. We consider a population of 1000 $people/km^2$ with 1 $person/km^2$ initially infected in all area of the domain. Then, the initial conditions are: $s_0=999$, $e_0 = 0$, $i_0=1$, $r_0=0$ and $d_0=0$. This test aims to reproduce a compartmental simulation of the EPIDEMIC software by using the same initial parameters. The results have to be the same in each point of the domain and the same as the EPIDEMIC software. We set $\alpha = 0.14286$ $days^{-1}$, $\beta = 0.25$ $days^{-1}$, $\delta = 0.06666$ $days^{-1}$, $\gamma = 0.1$ $days^{-1}$ and $\Delta t =1$ $day$. The mesh has $50\times 50$ bilinear quadrilateral elements. Figure \ref{fig:seird_epidemic} shows the comparison of the results, where we can see a very good agreement between both solutions.

\begin{figure}[htpb]
    \centering
    \includegraphics[width=0.6\linewidth]{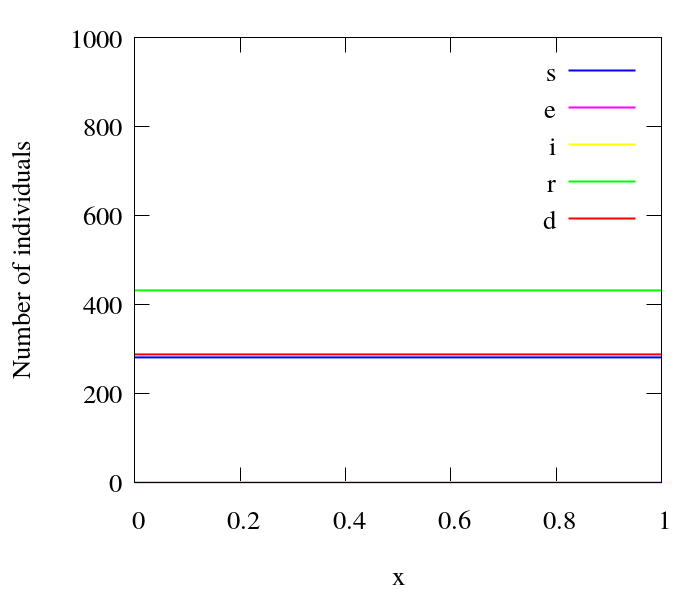}    
    \caption{Test 1: Values over a centralized horizontal line at t=365 days}
    \label{fig:line_test1}
\end{figure}

Figure \ref{fig:line_test1} shows the results over a centralized horizontal line crossing the domain at t=365 days. It is possible to see that the results are the same in all the domain, as expected.

\subsection{Test 2: Initial infected only in a circle region with diffusion}

\begin{figure}[htpb]
    \centering
    \includegraphics[width=0.6\linewidth]{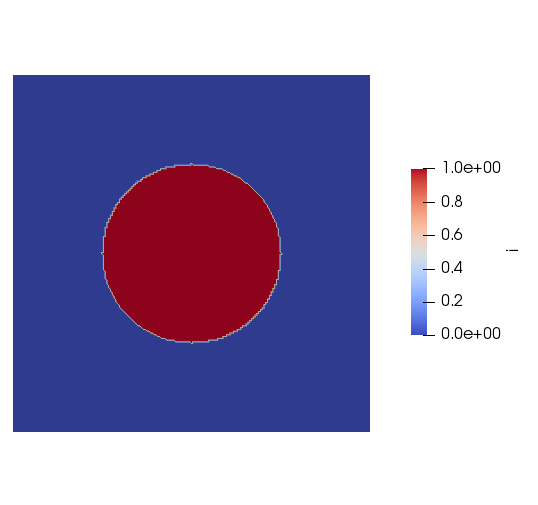}    
    \caption{Test 2: Infected initial condition}
    \label{fig:initial}
\end{figure}

Now, we consider the same parameters of the previous example, but different initial conditions. We consider a population of 1000 $people/km^2$ in all area of the domain with 1 $person/km^2$ initially infected only in a circle centered at $(0,0)$ and radius $R=0.5$ $km$, We assume that $\nu_s = \nu_e = \nu_i = \nu_r = 10^{-8}$ $km^2  persons^{-1}  days^{-1}$. Then, the initial conditions are: $s_0=999$, $e_0 = 0$, $i_0=1$ for $R<=0.5$ and $i_0=0$ for $R>0$ with $R=\sqrt{x^2+y^2}$, $r_0=0$ and $d_0=0$ (see Figure \ref{fig:initial}). We consider adaptive mesh refinement in this example. The original mesh has $50 \times 50$ bilinear quadrilateral elements, and after the refinement, the smallest element has size 0.005 $km$. We initially refine the domain in two levels. For the AMR/C procedure, we set $h_{max}=2$,$r_f=0.95$, $c_f=0.05$. We apply the adaptive mesh refinement every 5 time-steps. 




\begin{figure}[htpb]
    \centering
    \includegraphics[width=0.6\linewidth]{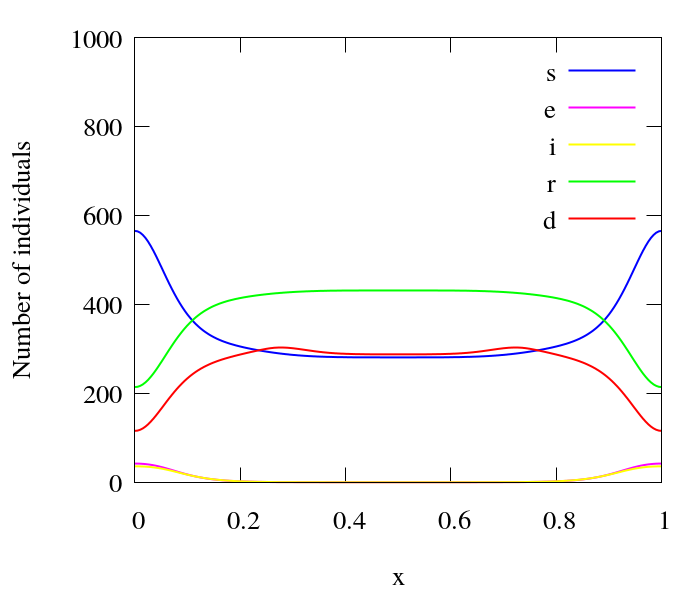}  
    \caption{Test 2: Values over a centralized horizontal line at t=365 days}
    \label{fig:line_test3}
\end{figure}

\begin{figure}[htpb]
    \centering
    \includegraphics[width=1\linewidth]{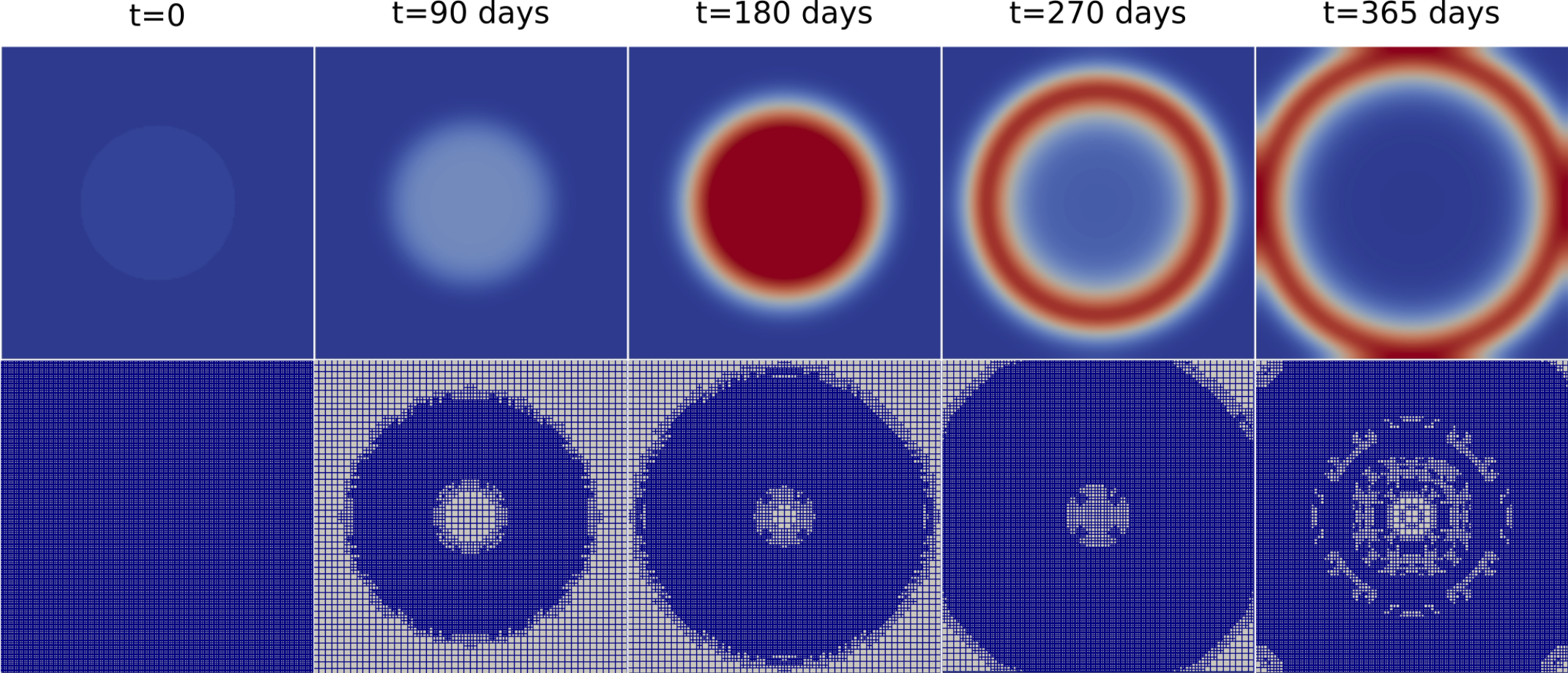}  
    \caption{Test 2: Infected people at different time-steps (top) and adapted meshes (bottom).}
    \label{fig:infected}
\end{figure}

Figure \ref{fig:line_test3} shows the results over a centralized horizontal line crossing the domain at t=365 days. Figure \ref{fig:infected} shows the infected people at different time-steps. 
Note that the infected remains actives in other parts of the domain because of the diffusion. It is possible to see the wave effect of the disease spreading. Note that the AMR/C procedure improves spatial resolution on the regions where the infected people are higher. 

\subsection{Test 3: Varying the population}

\begin{figure}[htpb]
    \centering
    \includegraphics[width=0.5\linewidth]{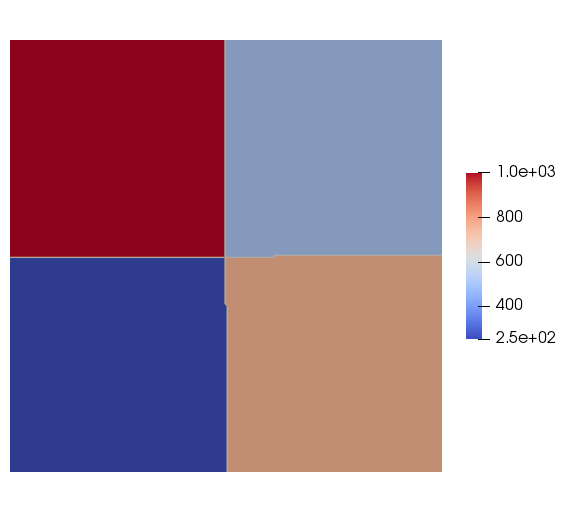}    
    \caption{Test 3: Susceptible initial condition}
    \label{fig:initial_s}
\end{figure}

In this test, we change the initial population. Instead of a constant value in all domain, we set 1000 $people/km^2$ at the left/top quadrant, 500 $people/km^2$ at the right/top quadrant, 250 $people/km^2$ at the left/bottom quadrant and 750 $people/km^2$ at the right/bottom quadrant (Figure \ref{fig:initial_s}). Then, the initial conditions are: $s_0=999$ for $x\leq0$ and $y>0$, $s_0=499$ for $x>0$ and $y>0$,  $s_0=249$ for $x\leq0$ and $y<=$,  $s_0=749$ for $x>0$ and $y>0$, $e_0 = 0$, $i_0=1$ for $R\leq0.5$ and $i_0=0$ for $R>0$ with $R=\sqrt{x^2+y^2}$, $r_0=0$ and $d_0=0$. The initial population infected is 1 $person/km^2$ at the same circled region of the previous test. All other parameters are the same of the previous simulation.

\begin{figure}[htpb]
    \centering
    \includegraphics[width=1\linewidth]{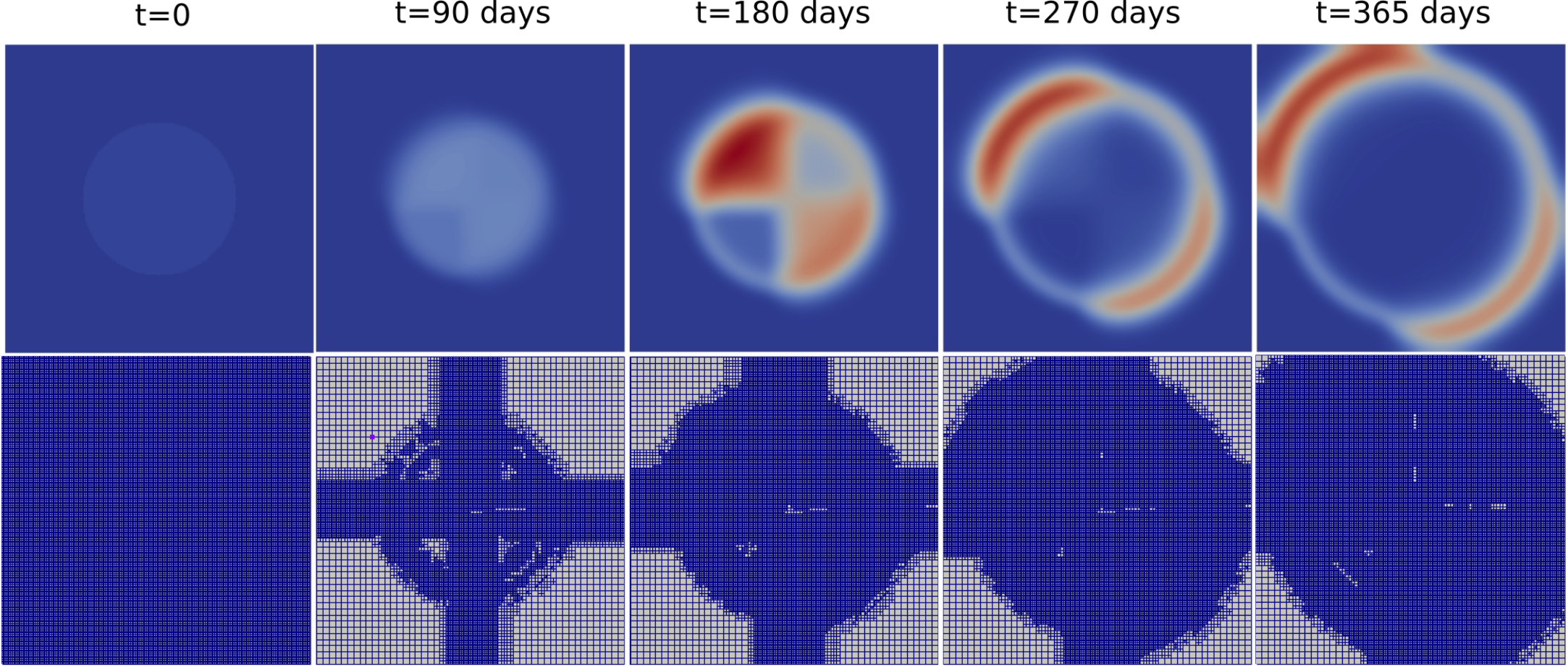}  
    \caption{Test 3: Infected people at different time-steps (top) and adapted meshes (bottom).}
    \label{fig:infected4}
\end{figure}

\begin{figure}[htpb]
    \centering
    \includegraphics[width=0.5\linewidth]{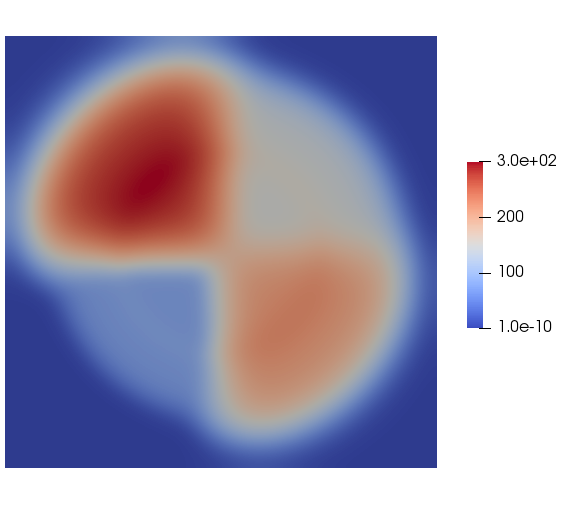}    
    \caption{Test 3: Total deaths at t=365 days.}
    \label{fig:deaths}
\end{figure}

Figure \ref{fig:infected4} shows the infected people ate different time-steps. It is possible to see that the regions with denser populations (more $people/km^2$) are more affected by the disease. Figure \ref{fig:deaths} shows the total number of deaths after 365 days, and the regions with more $people/km^2$ have more deaths than the less dense regions. Note also that the AMR/C procedure generates meshes following the model dynamics. 

\section{Numerical Results: Verification of the spatio-temporal model of COVID-19
infection spread}\label{NRC19}

In this section, we perform some simulations to validate the spatio-temporal model of COVID-19 infection spread.

\subsection{COVID19 Test 1: Compartmental model}

\begin{figure}[htpb]
    \centering
    \includegraphics[width=0.9\linewidth]{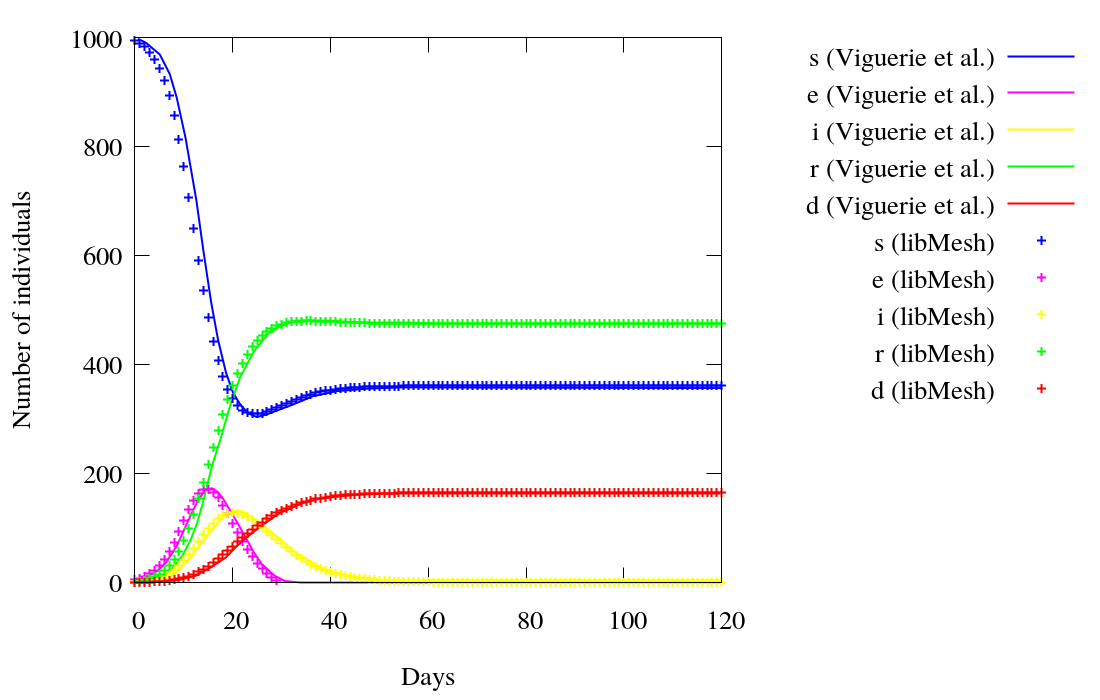}
    \caption{COVID19 Test 1: Compartmental model}
    \label{fig:seird_covid}
\end{figure}

In this test, we do not consider diffusion. We consider a square domain of $1km\times1km$ centered at $(0,0)$ with a population of 1000 $people/km^2$, with 1 $person/km^2$ initially infected and 5 $people/km^2$ exposed in all area of the domain.  Then, the initial conditions are: $s_0=994$, $e_0 = 5$, $i_0=1$, $r_0=0$ and $d_0=0$. The aim of this test is to reproduce a compartmental simulation presented in \cite{viguerie2020diffusion} by using the same initial parameters. The results has to be the same in each point of the domain and also the same of the ones given in \cite{viguerie2020diffusion}. We set $\alpha = 0.125$ $days^{-1}$, $\beta_i = \beta_e = 0.005 $ $days^{-1}persons^{-1}$, $\delta = 0.0625$ $days^{-1}$, $\gamma_i = 0.041666667$ $days^{-1}$ and $\gamma_e = 0.1666667$ $days^{-1}$. The mesh has $50\times 50$ bilinear quadrilateral elements. Figure \ref{fig:seird_covid} shows the comparison of the results, where we can see an excellent agreement.

\subsection{COVID19 Test 2: Reproducing a 1D model}

In this example, we reproduce a 1D model with quadrilateral elements being the spatial domain $\Omega$ given by $[0, 1]$ and a time interval $[0,T]$ with $T=200$ days. To reproduce a 1D simulation with quadrilateral elements, we fix the element width to 0.0005 and vary its length to find the proper refinement for this case. Therefore, we run a mesh convergence study as well as a time-step convergence study.

For the initial conditions, we set $s = s_0$ and $e = e_0$ as follows,

\begin{equation}
    s_0 = e^{-(x+1)^4}+e^{-\frac{(x-0.35) ^2}{10^{-2}}}+\frac{1}{8}\left(e^{-\frac{(x-0.62)^4}{10^{-5}}}+e^{-\frac{(x-0.52)^4}{10^{-5}}}+e^{-\frac{(x-0.42)^4}{10^{-5}}}\right) + \frac{1}{4}e^{-\frac{(x-0.735)^4}{10^{-5}}}
\end{equation}

\begin{equation}
    e_0=\frac{1}{20}e^{-\frac{(x-0.75)^4}{10^{-5}}}
\end{equation}

\begin{figure}[htpb]
    \centering
\includegraphics[width=0.6\linewidth]{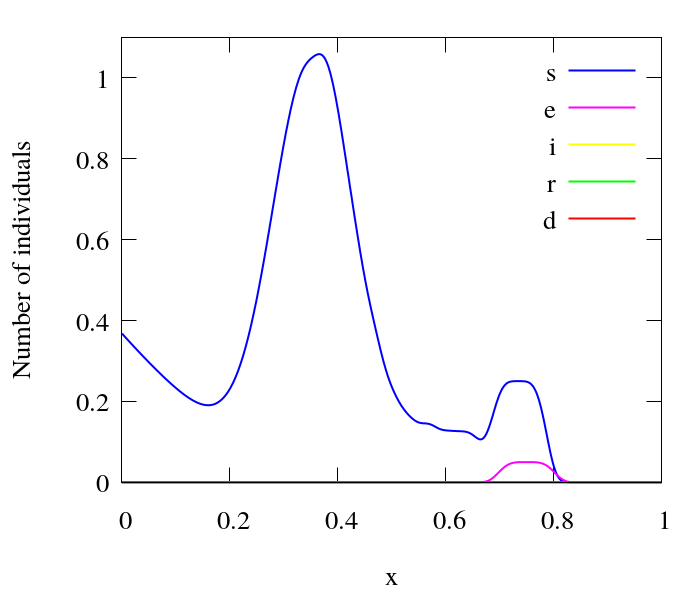}
    \caption{COVID19 Test 2: Initial conditions.}
    \label{fig:initial_conditions}
\end{figure}

Figure \ref{fig:initial_conditions} shows the initial conditions. We further set $i_0 = 0$, $r_0 = 0$, and $d_0 = 0$. Qualitatively, the initial conditions represent a large population centered around $x = 0.35$ with no exposed persons and a small population centered around $x = 0.75$ with some exposed individuals. We also enforce homogeneous Neumann boundary conditions at $x = 0$ and a zero-population Dirichlet boundary condition at $x = 1$ for all model compartments. The latter represents a non-populated area at $x = 1$.

 We set $\alpha = 0.09375$ $days^{-1}$, $\beta_i = \beta_e = 0.375$ $days^{-1}persons^{-1}$, $\delta = 0.0046875$ $days^{-1}$, $\gamma_i = 0.03125$ $days^{-1}$ and $\gamma_e = 0.125$ $days^{-1}$, $A=0$, $\nu_s = 3.75\times10^{-5}$, $\nu_e = 0.75\times 10^{-3}$, $\nu_i = 0.75\times10^{-10}$ and $\nu_r = 3.75\times 10^{-5}$ $km^2persons^{-1}days^{-1}$.
 
\begin{figure}[htpb]
    \centering
    \includegraphics[width=0.9\linewidth]{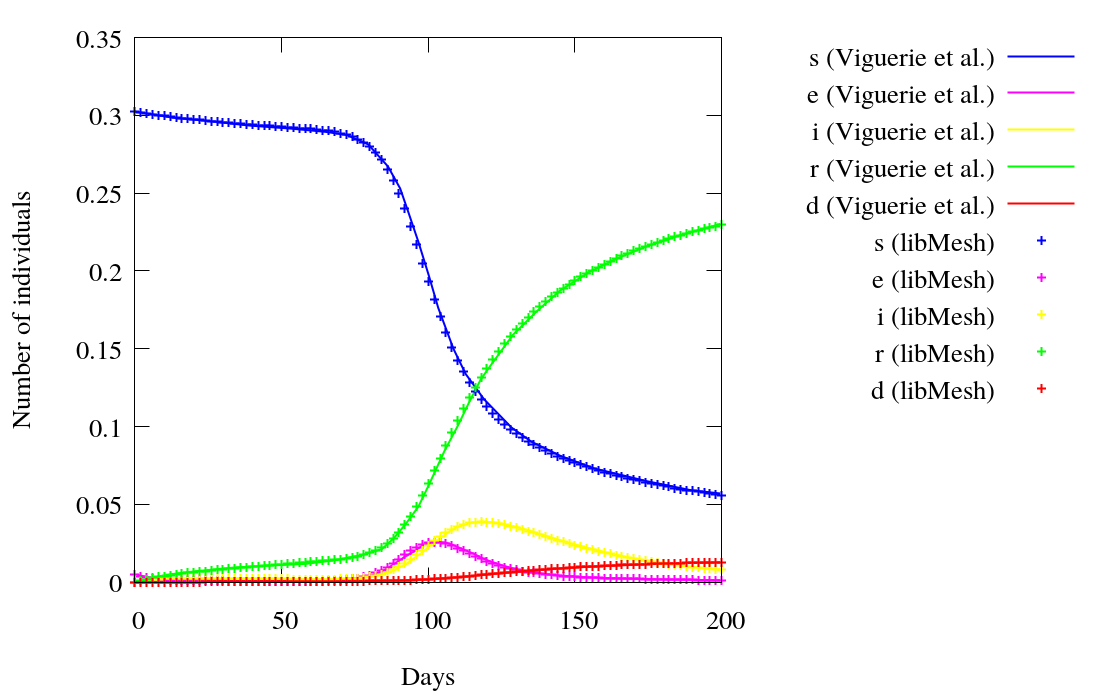}
    \caption{COVID19 Test 2: Reproducing a 1D model}
    \label{fig:seird_covid_diff}
\end{figure}

 Figure \ref{fig:seird_covid_diff} shows the comparison of the results with a mesh size $\Delta x = 1/500$ and a time-step $\Delta t = 0.25$ $days$. For comparison, we multiply the total number of individuals by 2000, since our element width is $1/2000$ and it has influence when integrating the domain. We can observe a very good agreement between both solutions.
 
\begin{figure}[htpb]
    \centering
    \includegraphics[width=0.6\textwidth]{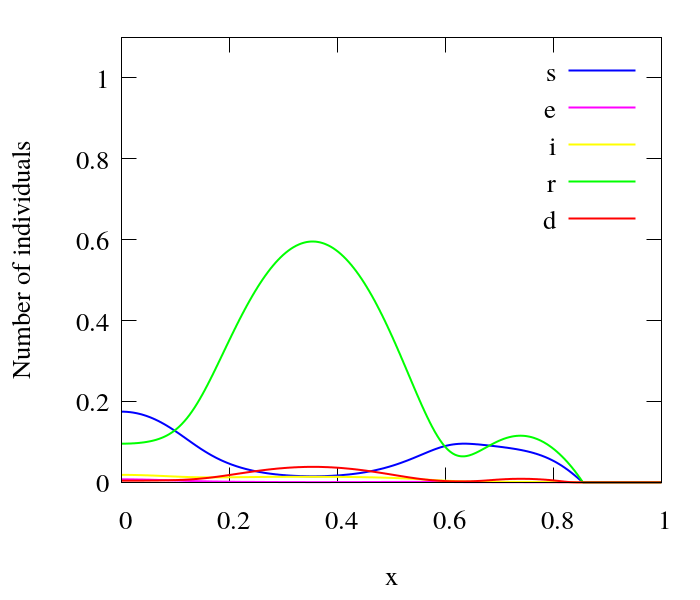}
    \caption{COVID Test 2: Populations at t=200 days.}
    \label{fig:1dresults}
\end{figure}

\subsubsection{Mesh convergence study}\label{mesh_conv}

\begin{figure}[htpb]
  \centering
  \subfloat{\includegraphics[width=0.5\textwidth]{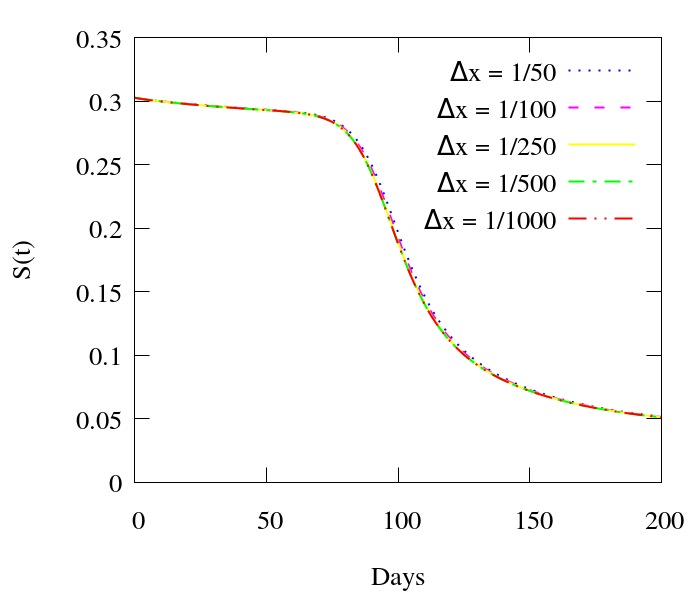}\label{s_mesh}}
  \hfill
  \subfloat{\includegraphics[width=0.5\textwidth]{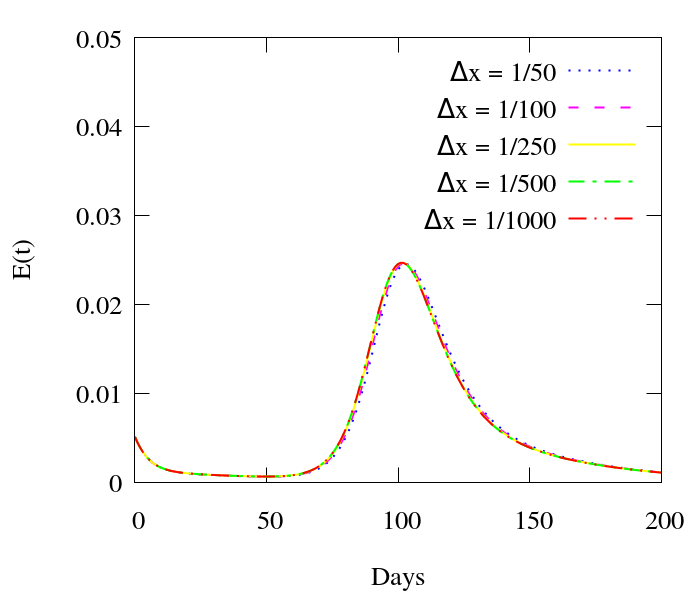}\label{e_mesh}}
    \hfill
  \subfloat{\includegraphics[width=0.5\textwidth]{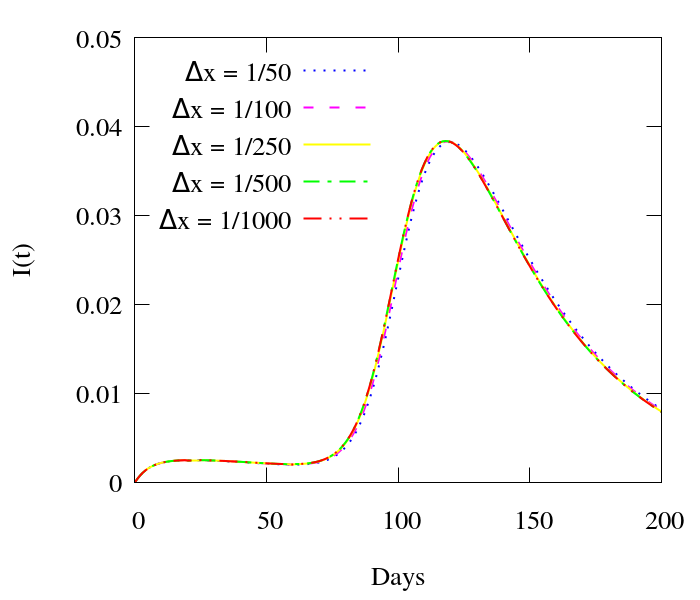}\label{i_mesh}}
    \hfill
  \subfloat{\includegraphics[width=0.5\textwidth]{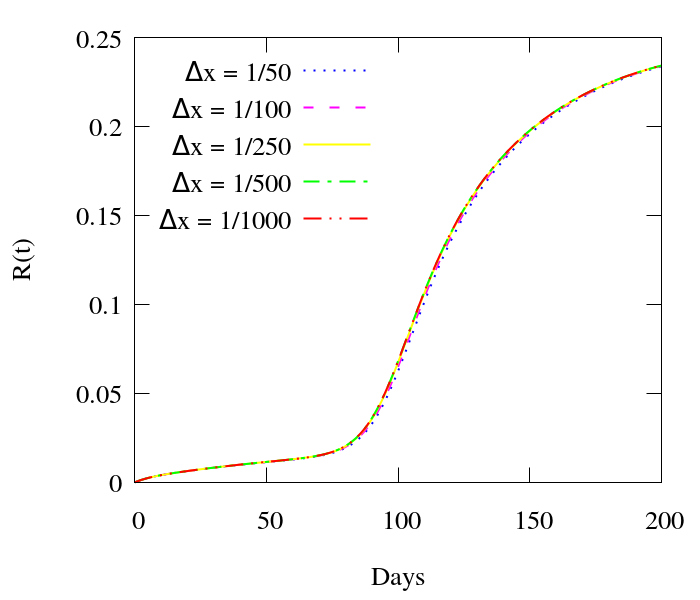}\label{r_mesh}}
      \hfill
  \subfloat{\includegraphics[width=0.5\textwidth]{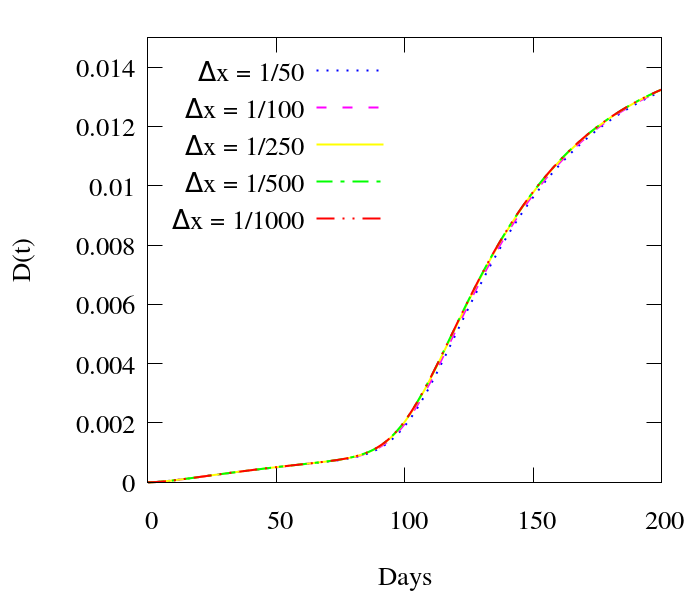}\label{d_mesh}}
  \caption{COVID19 Test 2: Mesh convergence study (Total population by time).}
  \label{mesh_time}
\end{figure}

We compare numerical solutions computed on successively refined uniform grids with mesh size $\Delta x=1/50, 1/100, 1/250, 1/500$, and $1/1000$. The time step is $\Delta t = 0.25$ days. Figure \ref{mesh_time} shows the difference in the total population of each compartment of individuals for the different meshes.

\begin{figure}[htpb]
  \centering
  \subfloat{\includegraphics[width=0.5\textwidth]{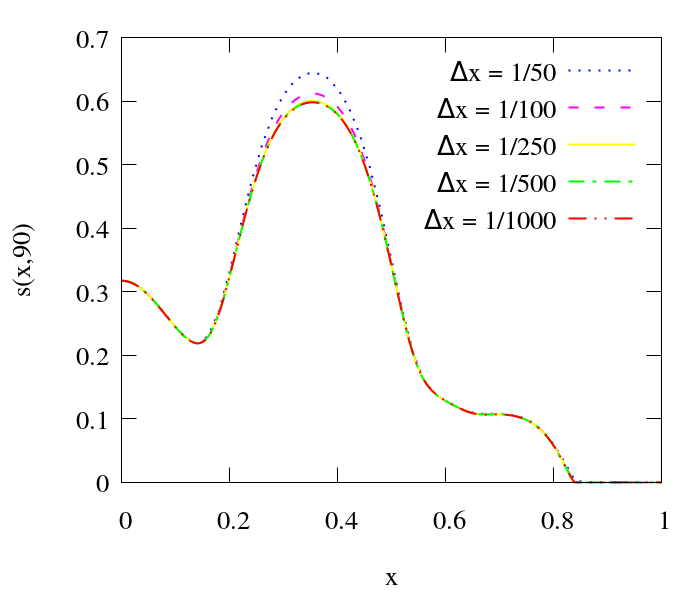}\label{s_mesh90}}
  \hfill
  \subfloat{\includegraphics[width=0.5\textwidth]{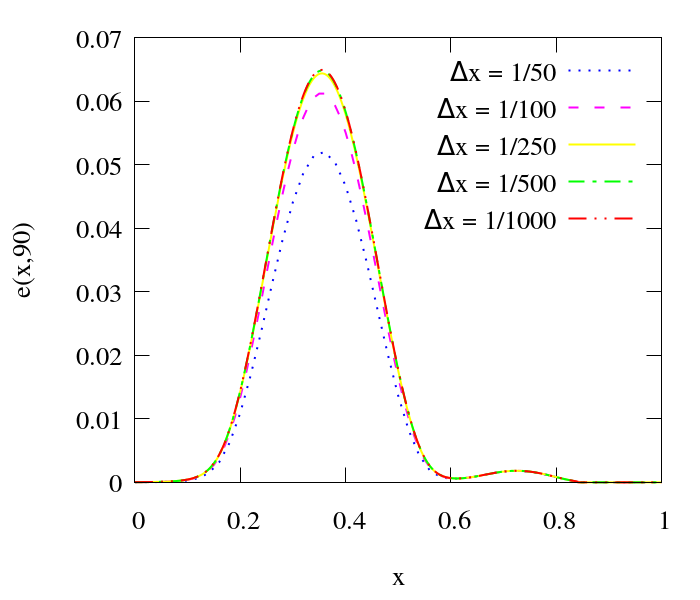}\label{e_mesh90}}
    \hfill
  \subfloat{\includegraphics[width=0.5\textwidth]{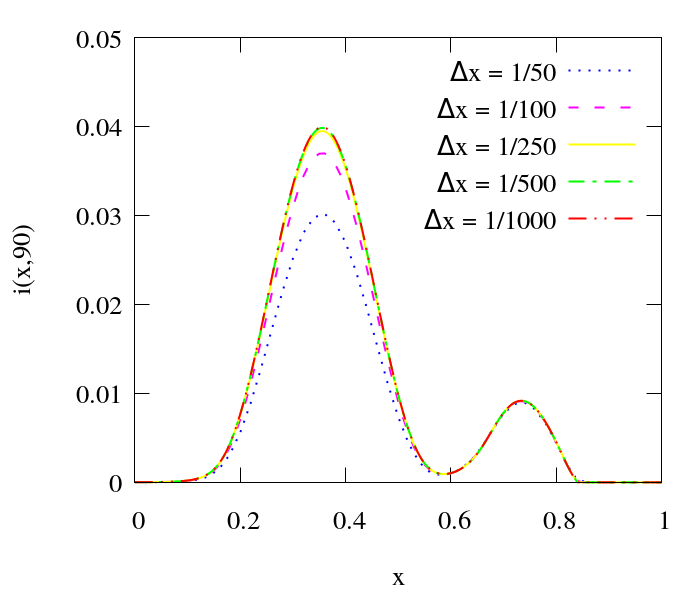}\label{i_mesh90}}
    \hfill
  \subfloat{\includegraphics[width=0.5\textwidth]{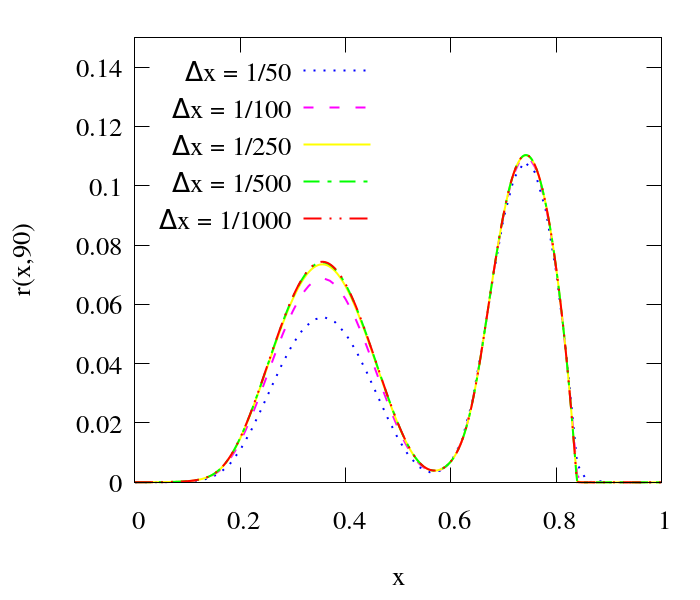}\label{r_mesh90}}
      \hfill
  \subfloat{\includegraphics[width=0.5\textwidth]{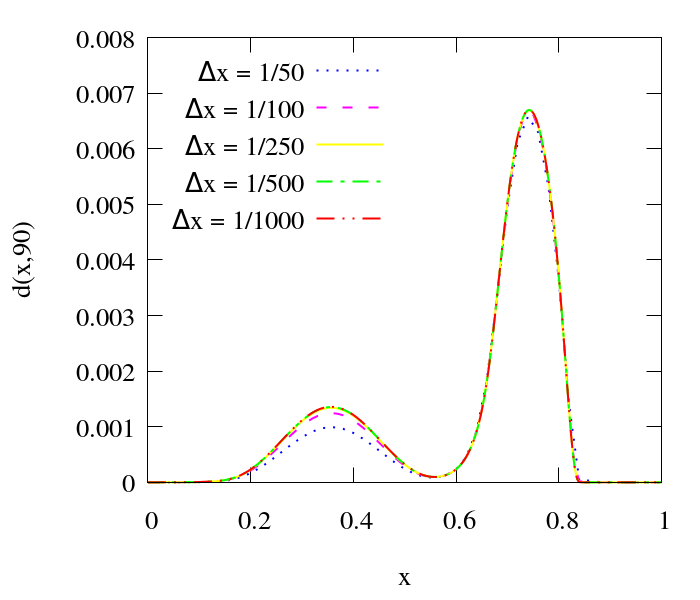}\label{d_mesh90}}
  \caption{COVID19 Test 2: Mesh convergence study (Individuals at t=90 days).}
  \label{mesh_spatial}
\end{figure}

A good resolution is found for $\Delta x=1/500$. It is easy to see this convergence in Figure \ref{mesh_spatial}, where the number of individuals of each compartment is plotted at $t=90$  days.

\subsubsection{Time-step convergence study}

\begin{figure}[htpb]
  \centering
  \subfloat{\includegraphics[width=0.5\textwidth]{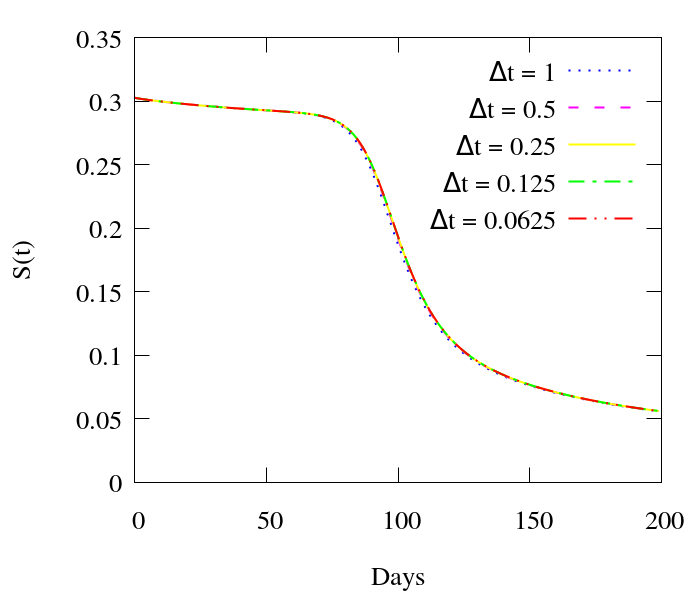}\label{s_time}}
  \hfill
  \subfloat{\includegraphics[width=0.5\textwidth]{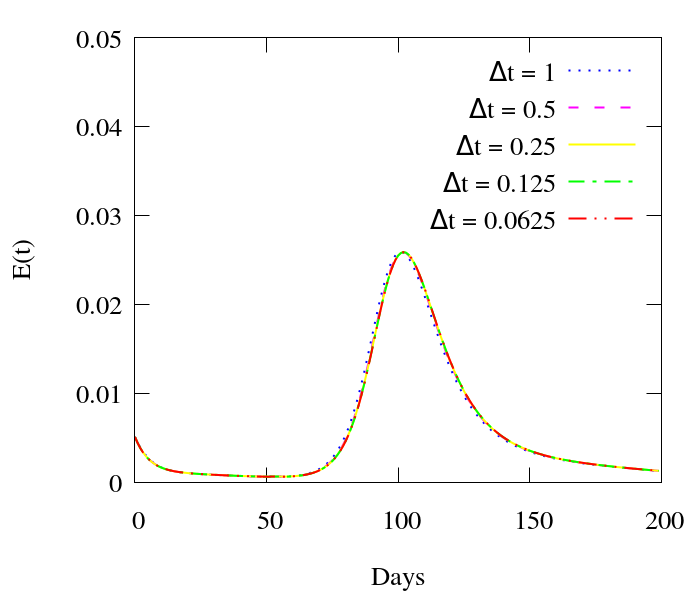}\label{e_time}}
    \hfill
  \subfloat{\includegraphics[width=0.5\textwidth]{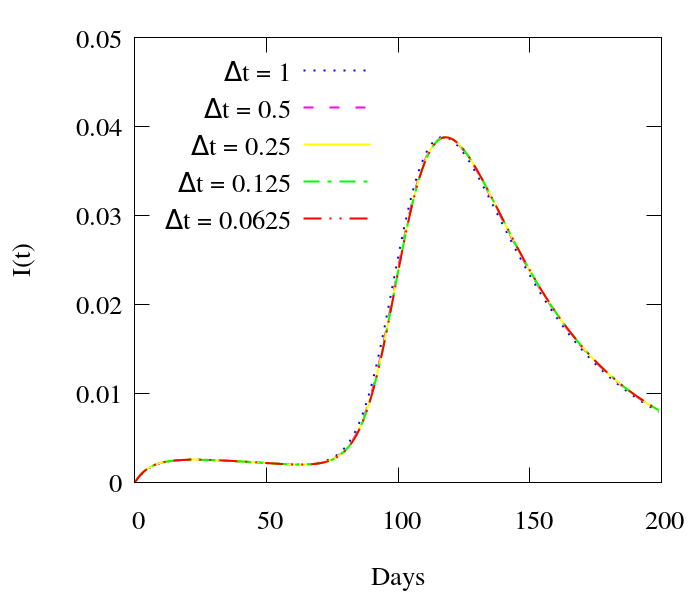}\label{i_time}}
    \hfill
  \subfloat{\includegraphics[width=0.5\textwidth]{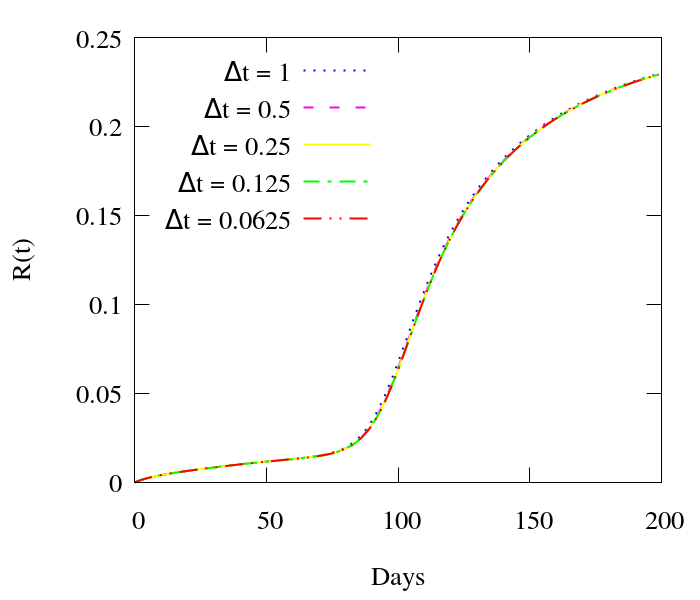}\label{r_time}}
      \hfill
  \subfloat{\includegraphics[width=0.5\textwidth]{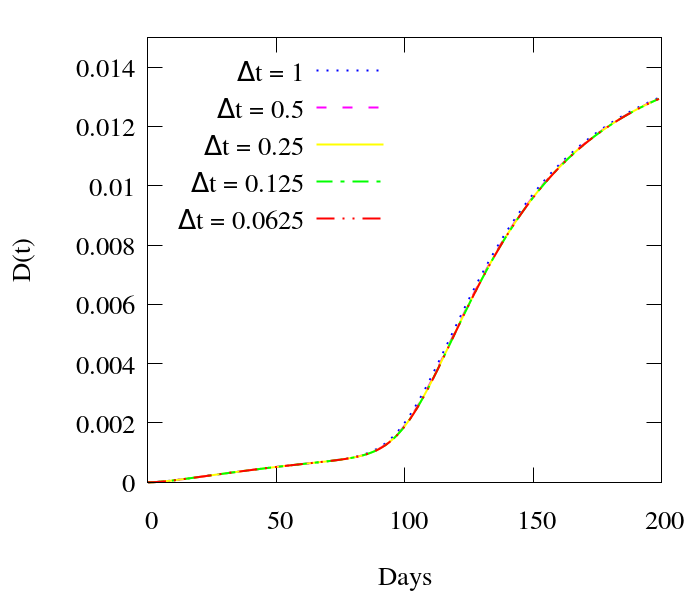}\label{d_time}}
  \caption{COVID19 Test 2: Time convergence study (Total population by time).}
  \label{time_time}
\end{figure}

We examine the impact of time-step size $\Delta t$ on the numerical approximation of the model solution. We consider the time step sizes $\Delta t = 1$, $\Delta t = 0.5$, $\Delta t = 0.25$, $\Delta t = 0.125$ and $\Delta t = 0.0625$ days. As the results in Section \ref{mesh_conv} suggested $\Delta x =1/500$ is a sufficiently fine spatial discretization, we utilize this mesh resolution here. Figure \ref{time_time} shows the difference of the total population of each compartment of individuals for the different time-steps.

\begin{figure}[htpb]
  \centering
  \subfloat{\includegraphics[width=0.5\textwidth]{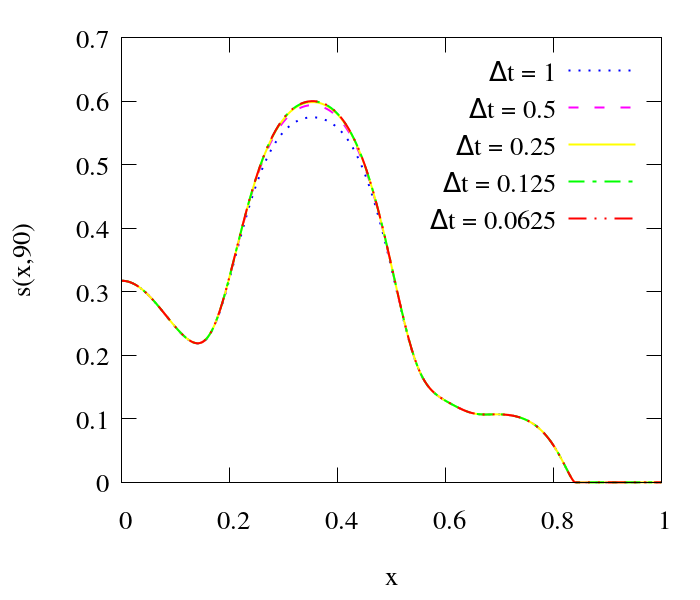}\label{s_time90}}
  \hfill
  \subfloat{\includegraphics[width=0.5\textwidth]{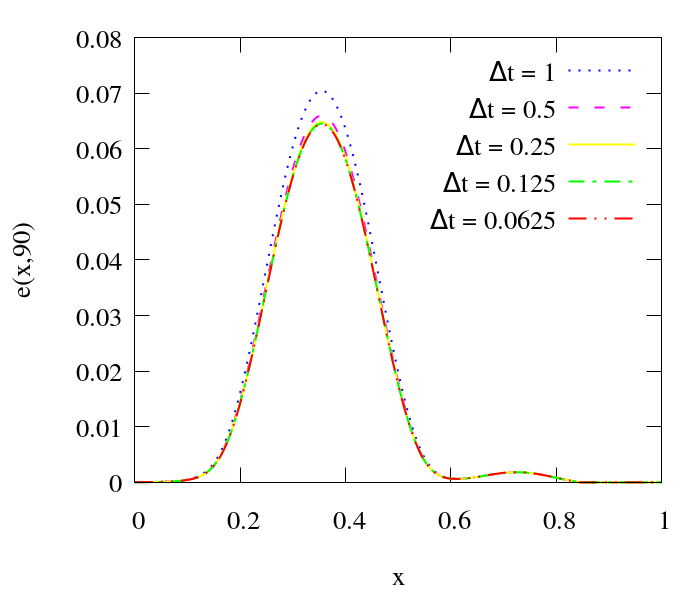}\label{e_time90}}
    \hfill
  \subfloat{\includegraphics[width=0.5\textwidth]{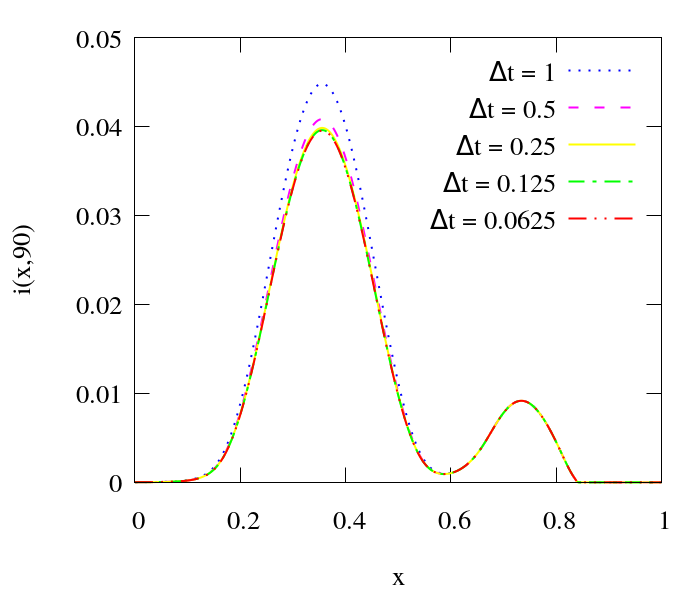}\label{i_time90}}
    \hfill
  \subfloat{\includegraphics[width=0.5\textwidth]{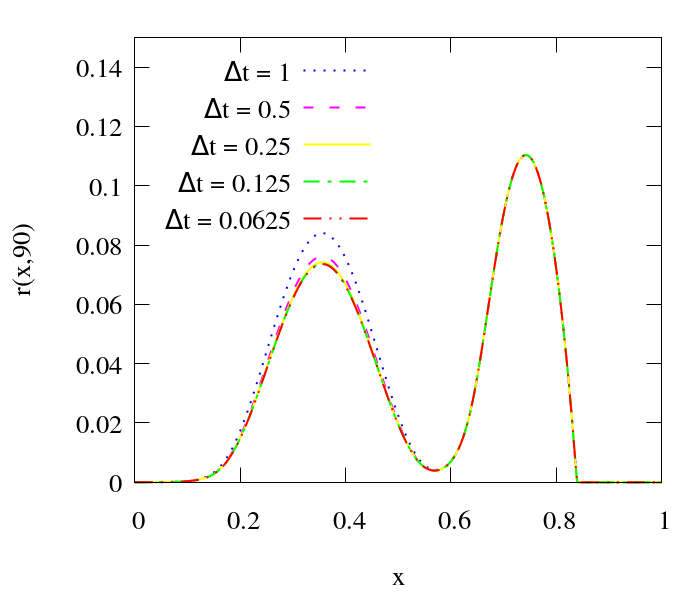}\label{r_time90}}
      \hfill
  \subfloat{\includegraphics[width=0.5\textwidth]{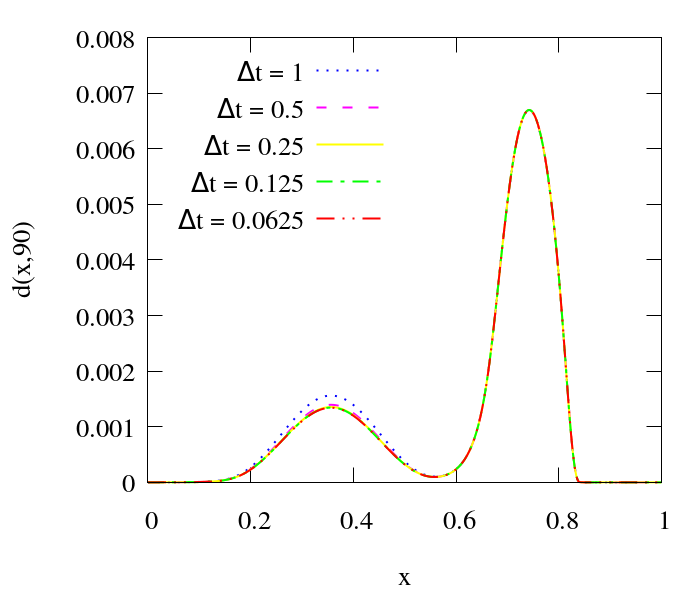}\label{d_time90}}
  \caption{COVID19 Test 2: Time convergence study (Individuals at t=90 days).}
  \label{time_spatial}
\end{figure}

A good accuracy is found for $\Delta t=0.25$ days. It is easy to see how the accuracy improves in Figure \ref{time_spatial}, where the number of individuals of each compartment is plotted at $t=90$  days.

\subsection{COVID19 Test 3: Reproducing a 2D model}

This test is the application of the previous configuration rotated in a two dimensional square with corners at (-1,-1), (1,-1), (1,1) and (-1,1). The initial population is:

\begin{equation}
    s_0 = e^{-(R+1)^4}+e^{-\frac{(R-0.35) ^2}{10^{-2}}}+\frac{1}{8}\left(e^{-\frac{(R-0.62)^4}{10^{-5}}}+e^{-\frac{(R-0.52)^4}{10^{-5}}}+e^{-\frac{(R-0.42)^4}{10^{-5}}}\right) + \frac{1}{4}e^{-\frac{(R-0.735)^4}{10^{-5}}}
\end{equation}

\begin{equation}
    e_0=\frac{1}{20}e^{-\frac{(R-0.75)^4}{10^{-5}}}
\end{equation}

\noindent with $R=\sqrt{x^2+y^2}$.

\begin{figure}[htpb]
    \centering
    \includegraphics[width=\textwidth]{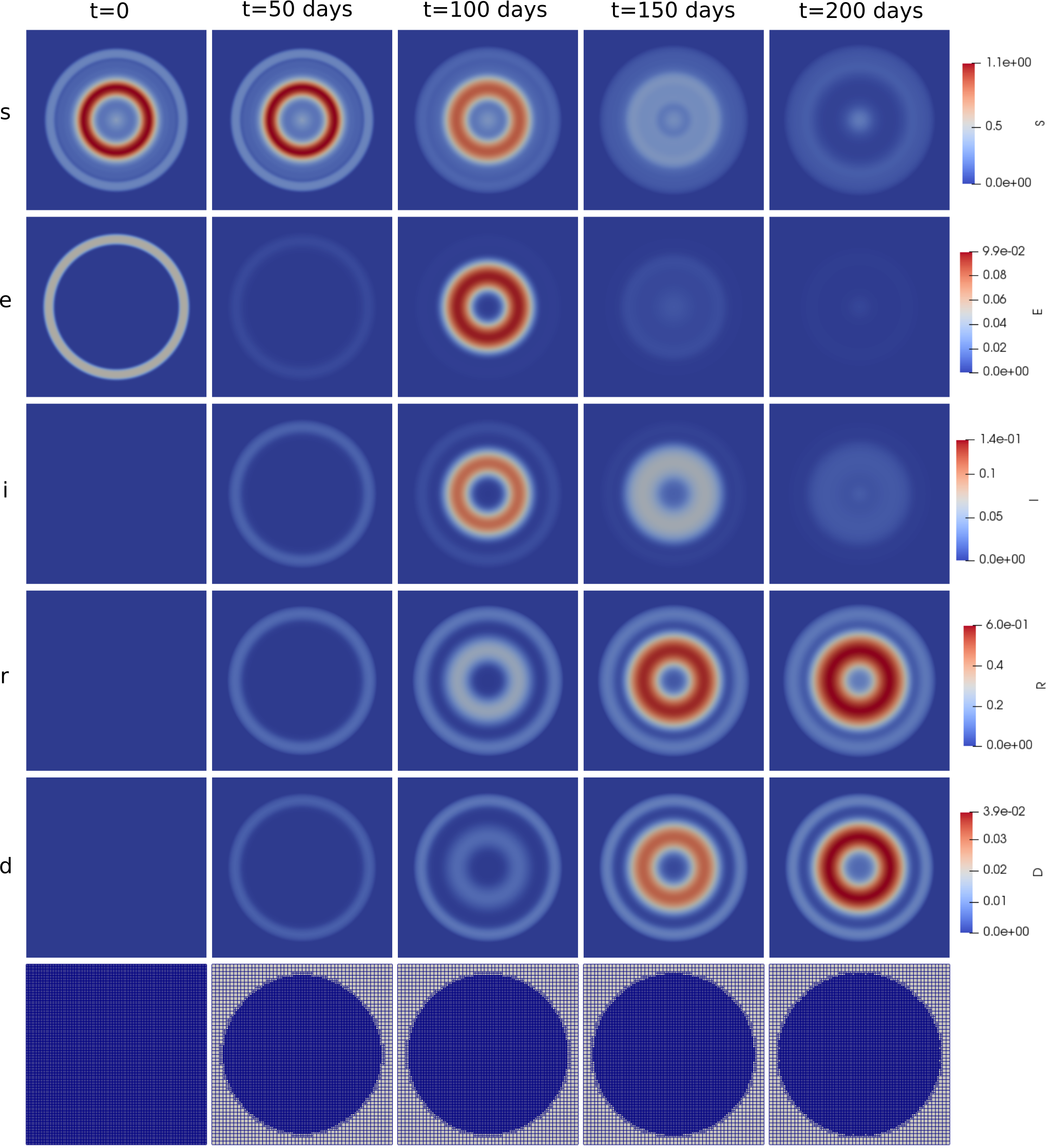}
    \caption{COVID Test 3: Populations at different times (top rows) and adapted meshes (bottom).}
    \label{fig:seird2d}
\end{figure}

 The original mesh has $50\times50$ bilinear quadrilaterals elements and it is refined in two levels at the beginning of the simulation. For the AMR/C procedure, we set $h_{max}=2$, $r_f=0.95$, $c_f=0.05$. We apply the adaptive mesh refinement every 4 time-steps. The behavior of the transmission has to be similar to the 1D model results, but in a radial configuration. Figures \ref{fig:seird2d} shows the populations at different time steps. Figure \ref{fig:line_test3} shows the results over a centralized horizontal line (or vertical because the axisymmetry) crossing the domain at t=200 days. If we compare Figure \ref{fig:line_test3} with Figure \ref{fig:1dresults}, it is possible to see that the populations follow a similar behavior.

\begin{figure}[htpb]
    \centering
    \includegraphics[width=0.9\textwidth]{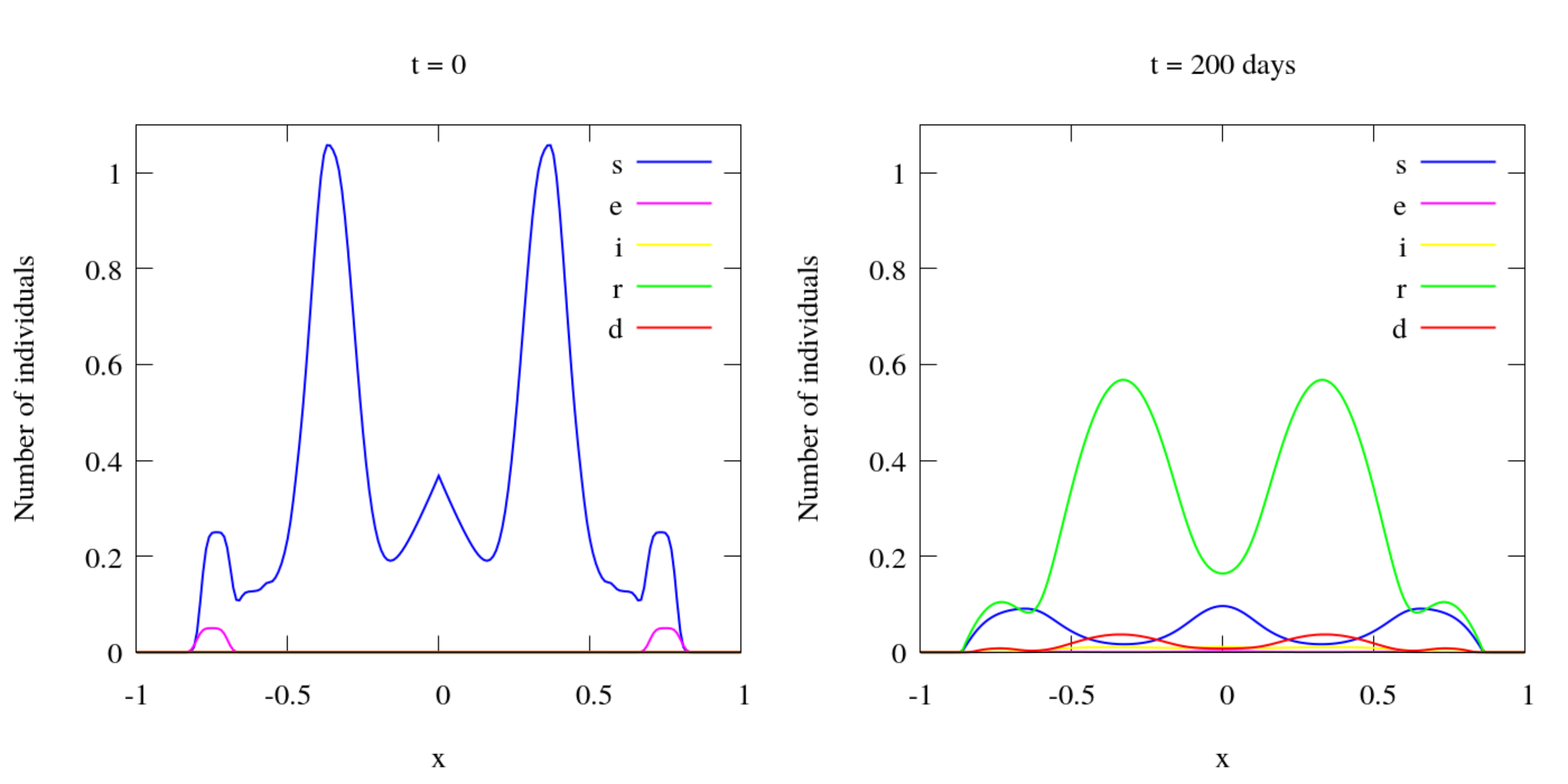}
    \caption{COVID Test 3: Populations over a horizontal/vertical line crossing the middle of the domain.}
    \label{fig:xx-iso}
\end{figure}

\begin{figure}[htpb]
    \centering
    \includegraphics[width=0.5\textwidth]{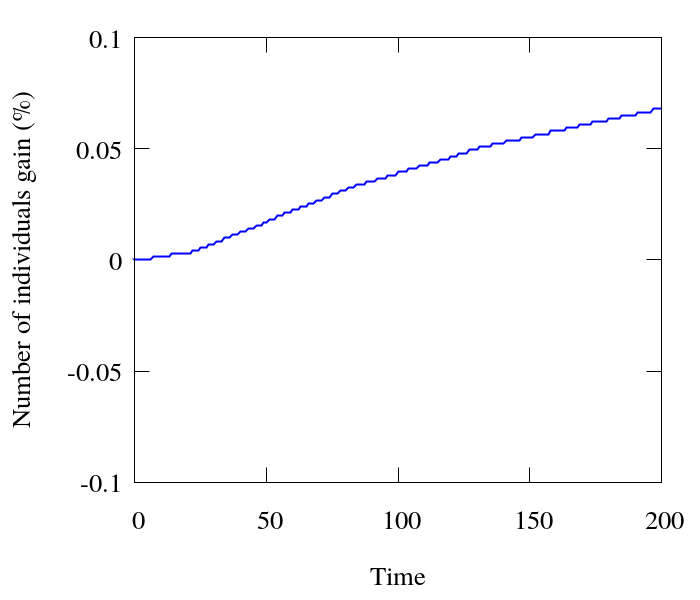}
    \caption{COVID Test 3: Time history of the total number of individuals.}
    \label{fig:mass_iso}
\end{figure}

In Figure \ref{fig:mass_iso} we plot the time history of the total number of individuals. There is a small gain in the total number of individuals (less than 0.1\%). 

\subsection{COVID19 Test 4: Anisotropic diffusion}

\begin{figure}[htpb]
    \centering
    \includegraphics[width=0.9\textwidth]{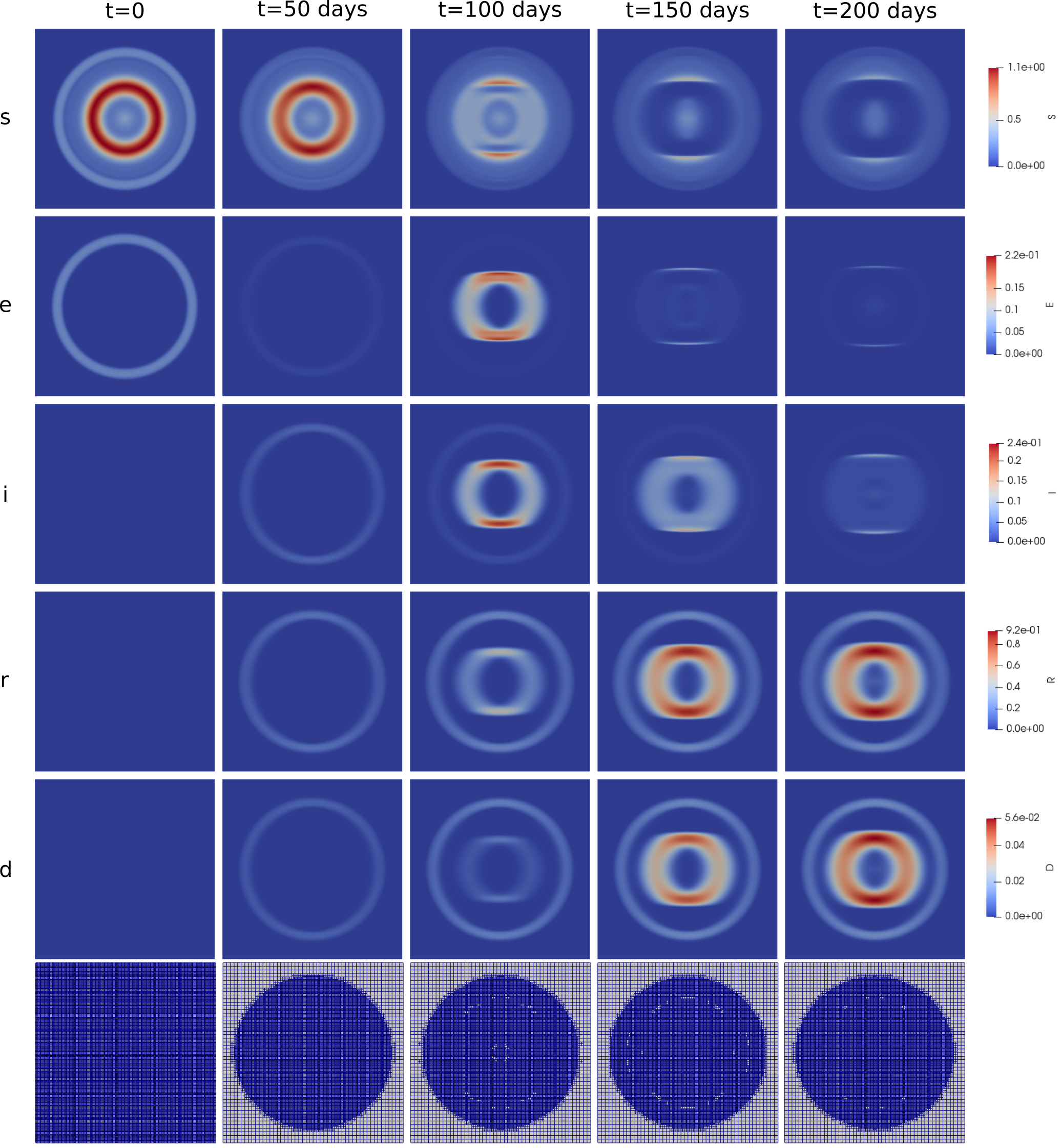}
    \caption{COVID Test 4: Populations at different times (top rows) and adapted meshes (bottom)}
    \label{fig:diff_aniso}
\end{figure}

\begin{figure}[htpb]
    \centering
    \includegraphics[width=0.9\textwidth]{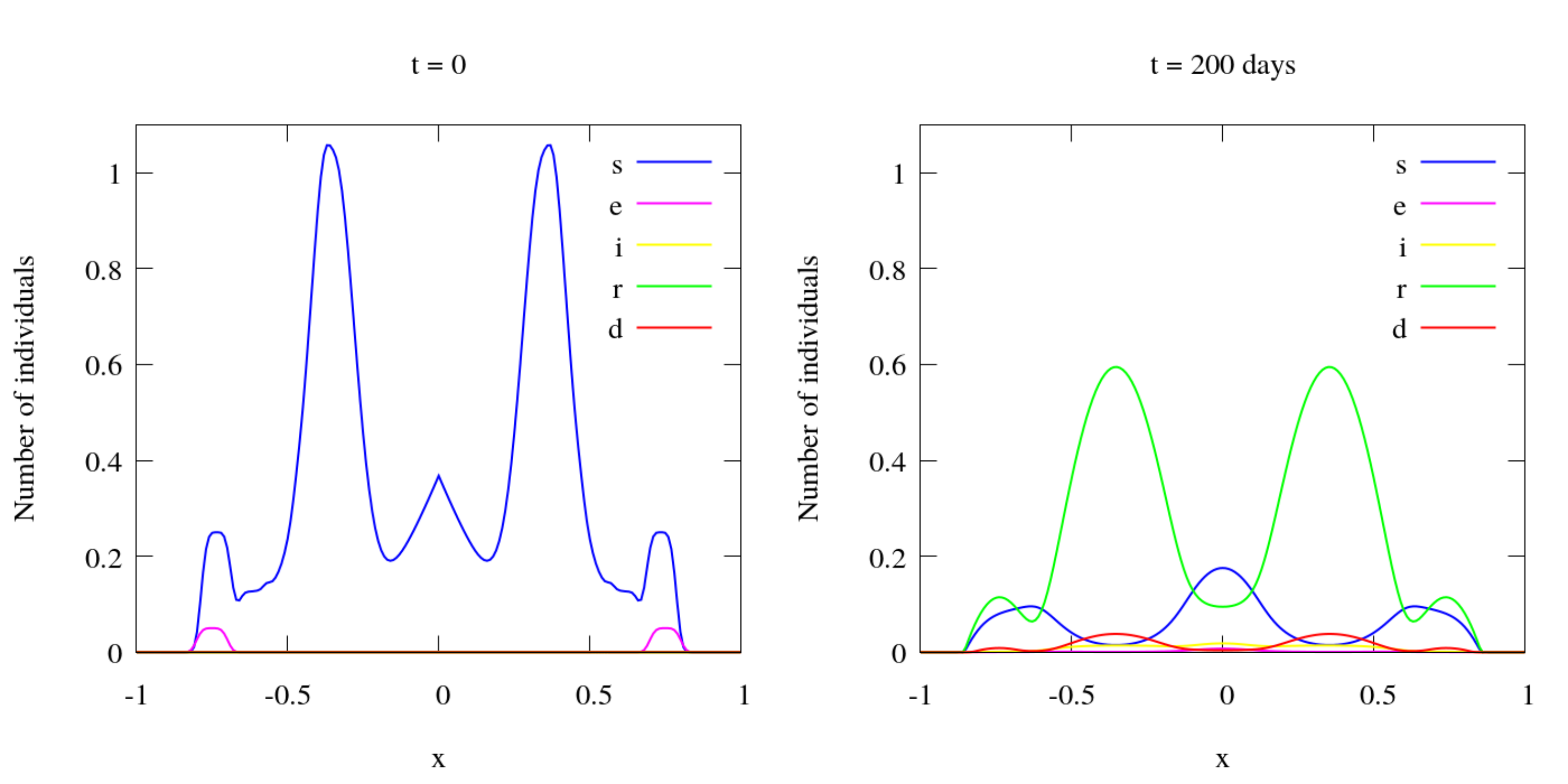}
    \caption{COVID Test 4: Populations over a horizontal line crossing the middle of the domain.}
    \label{fig:xx}
\end{figure}

\begin{figure}[htpb]
    \centering
    \includegraphics[width=0.9\textwidth]{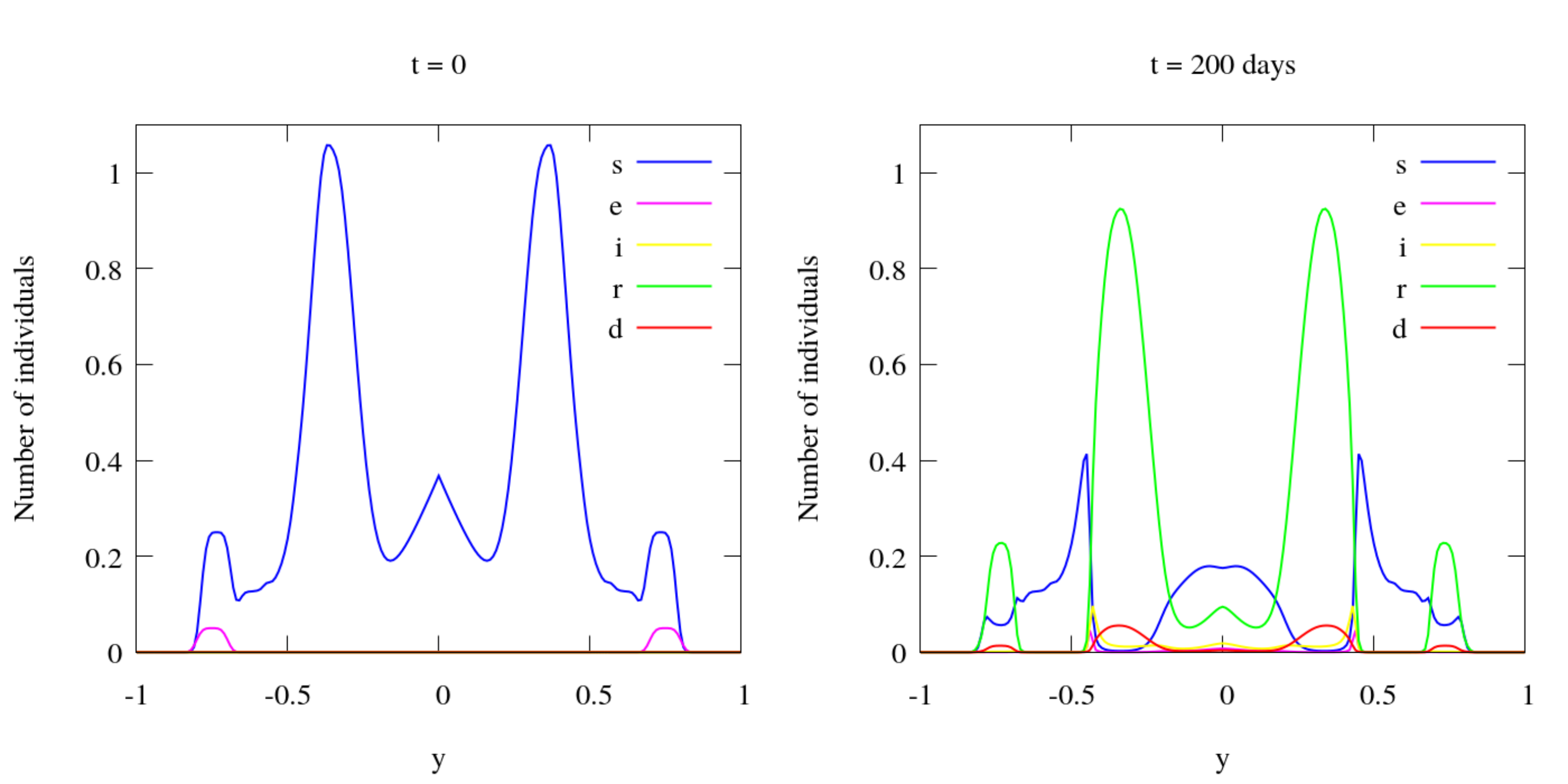}
    \caption{COVID Test 4: Populations over a vertical line crossing the middle of the domain.}
    \label{fig:yy}
\end{figure}

This test considers anisotropic diffusion in the previous configuration (only in the $x$ direction). Therefore, the populations move spatially only in the $x$ direction.  Figure \ref{fig:diff_aniso} shows the populations at different time-steps. Figure \ref{fig:xx} shows the results over a centralized horizontal line crossing the domain, and Figure \ref{fig:yy} over a centralized vertical line. By comparing these two figures, it is clear how the diffusion direction influences the behavior of the virus spread. Since there is no movement of infected or exposed people in the $y$ direction, part of the population does not have contact with the virus because there is no chance of the virus to reach them.

\begin{figure}[htpb]
    \centering
    \includegraphics[width=0.5\textwidth]{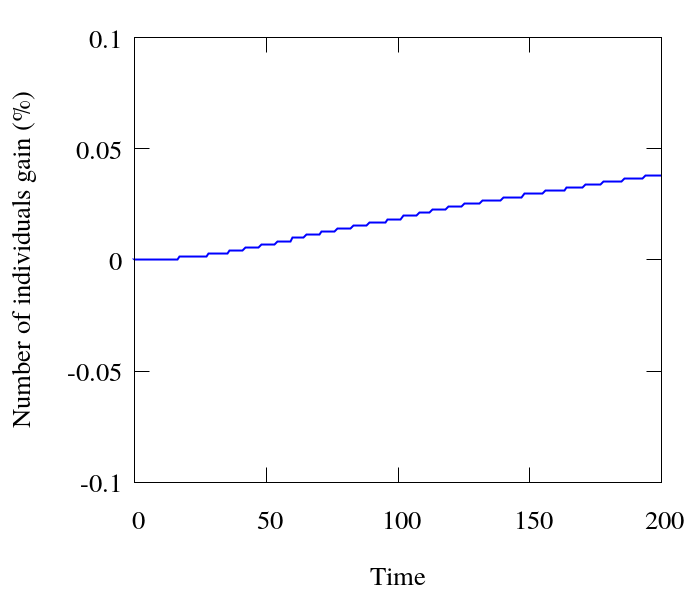}
    \caption{COVID Test 4: Time history of the total number of individuals.}
    \label{fig:mass_aniso}
\end{figure}

In Figure \ref{fig:mass_aniso} we plot the time history of the total number of individuals. We can see a gain in the total number of individuals of less than 0.1\%. 

\subsection{COVID19 Test 5: Random source}

This test has a new configuration. We still work with the two dimensional square with corners at (-1,-1), (1,-1), (1,1) and (-1,1) and an anisotropic diffusion only in the $x$ direction. We set $\alpha = 0.09375$ $days^{-1}$, $\beta_i = \beta_e = 0.375 /n$ $days^{-1}persons^{-1}$, $\delta = 0.0046875$ $days^{-1}$, $\gamma_i = 0.03125$ $days^{-1}$ and $\gamma_e = 0.125$ $days^{-1}$, $A=0$, $\nu_s = 3.75\times10^{-9}$, $\nu_e = 0.75\times 10^{-7}$, $\nu_i = 0.75\times10^{-14}$ and $\nu_r = 3.75\times 10^{-9}$ $km^2persons^{-1}days^{-1}$, and $\Delta t=0.25$ $days$.
 
 The original mesh has $50\times50$ bilinear quadrilaterals elements and it is refined in two levels at the beginning of the simulation. For the AMR/C procedure, we set $h_{max}=2$, $r_f=0.95$, $c_f=0.05$. We apply the adaptive mesh refinement every 4 time-steps.

The initial population is:

\begin{equation}
\begin{split}
s_0 = max \left\{\begin{array}{ccc}
100000e^{-\frac{R_1^4}{10^{-2}}}
\\
10000e^{-\frac{R_2^4}{10^{-4}}}
\\
10000e^{-\frac{R_3^4}{10^{-4}}}
\\
1000
\end{array}\right.
\end{split}
\end{equation}

\begin{equation}
    e_0 = 0 
\end{equation}

\begin{equation}
    i_0 = 0 
\end{equation}

\begin{equation}
    R_1 = \sqrt{x^2+y^2} 
\end{equation}

\begin{equation}
    R_2 = \sqrt{x^2+(y-0.75)^2} 
\end{equation}

\begin{equation}
    R_3 = \sqrt{(x-0.75)^2+y^2} 
\end{equation}

\begin{figure}[htpb]
    \centering
    \includegraphics[width=0.6\textwidth]{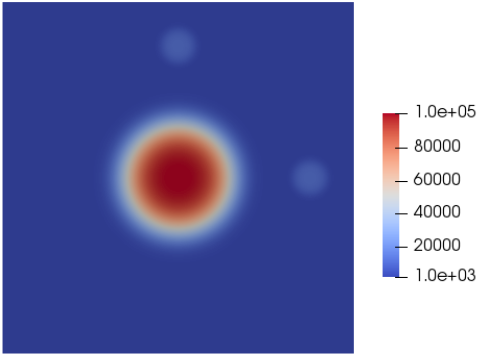}
    \caption{COVID Test 5: Initial susceptible population.}
    \label{fig:inital-source}
\end{figure}

\begin{figure}[htpb]
    \centering
    \includegraphics[width=0.5\textwidth]{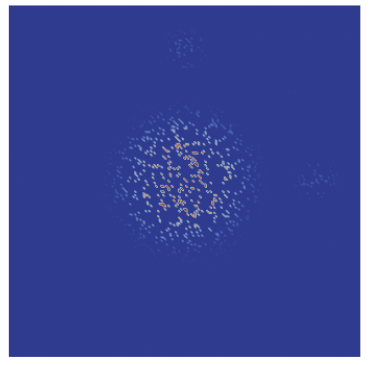}
    \caption{COVID Test 5: Example of the random source of exposed people.}
    \label{fig:random-source}
\end{figure}

Figure \ref{fig:inital-source} shows the initial susceptible population. Note there are not infected or exposed people at the initial time. We implement a random source of the exposed population that depends on the number of susceptible people. In all time-steps random nodes of the domain receive a certain number of exposed people. It tries to simulate people who travel and suddenly appear in a region carrying the virus. The random source does not add individuals to the population, but change individuals between susceptible and exposed compartments. Of course, this model is simple. Nevertheless, it demonstrates how to handle a random source term in the equations. Figure \ref{fig:random-source} shows a example of the random exposed number of people that appears in one time-step.

\begin{figure}[htpb]
    \centering
    \includegraphics[width=\textwidth]{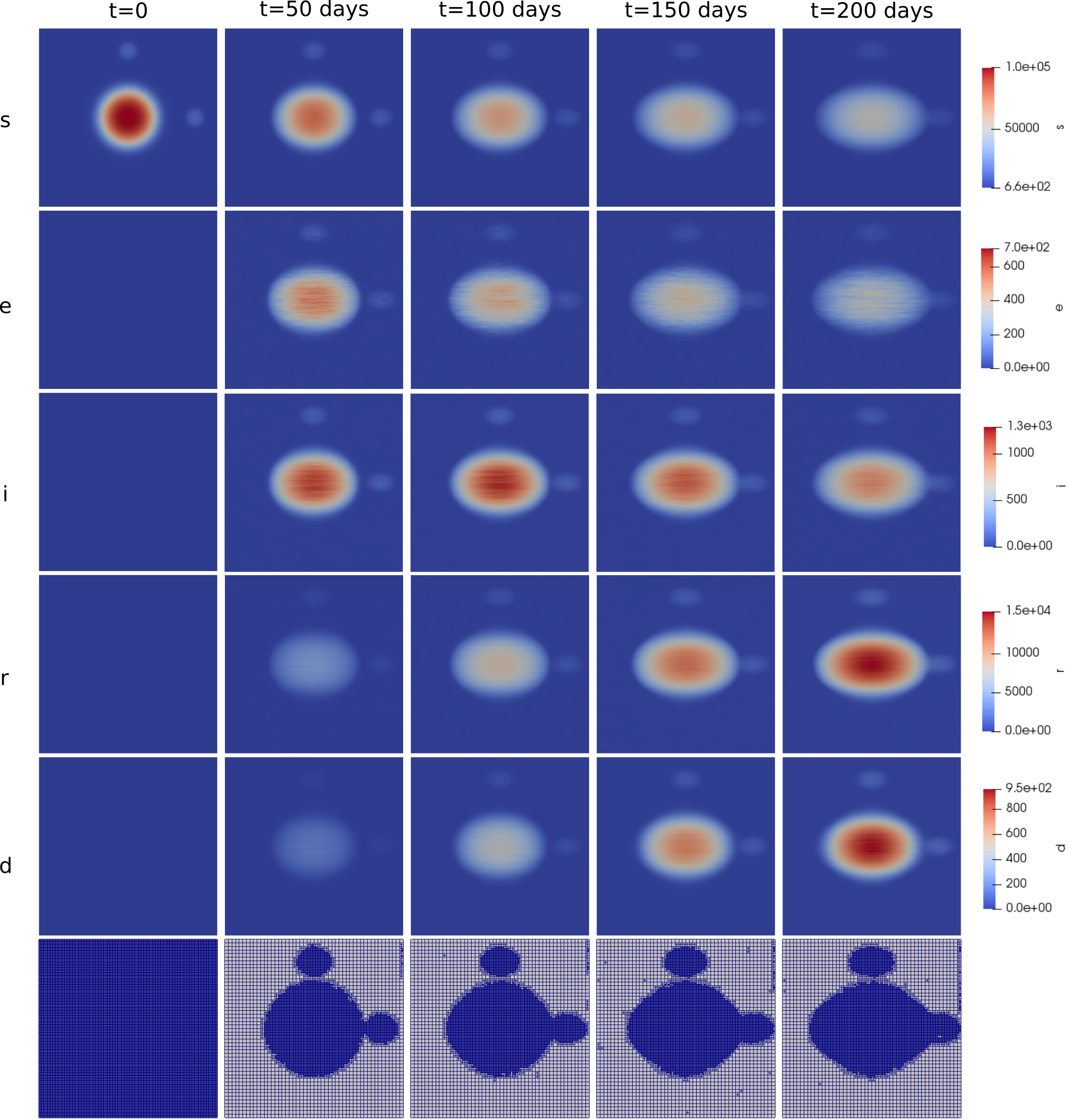}
    \caption{COVID Test 5: Populations at different times (top rows) and adapted meshes (bottom).}
    \label{fig:source-time}
\end{figure}

\begin{figure}[htpb]
    \centering
    \includegraphics[width=0.9\textwidth]{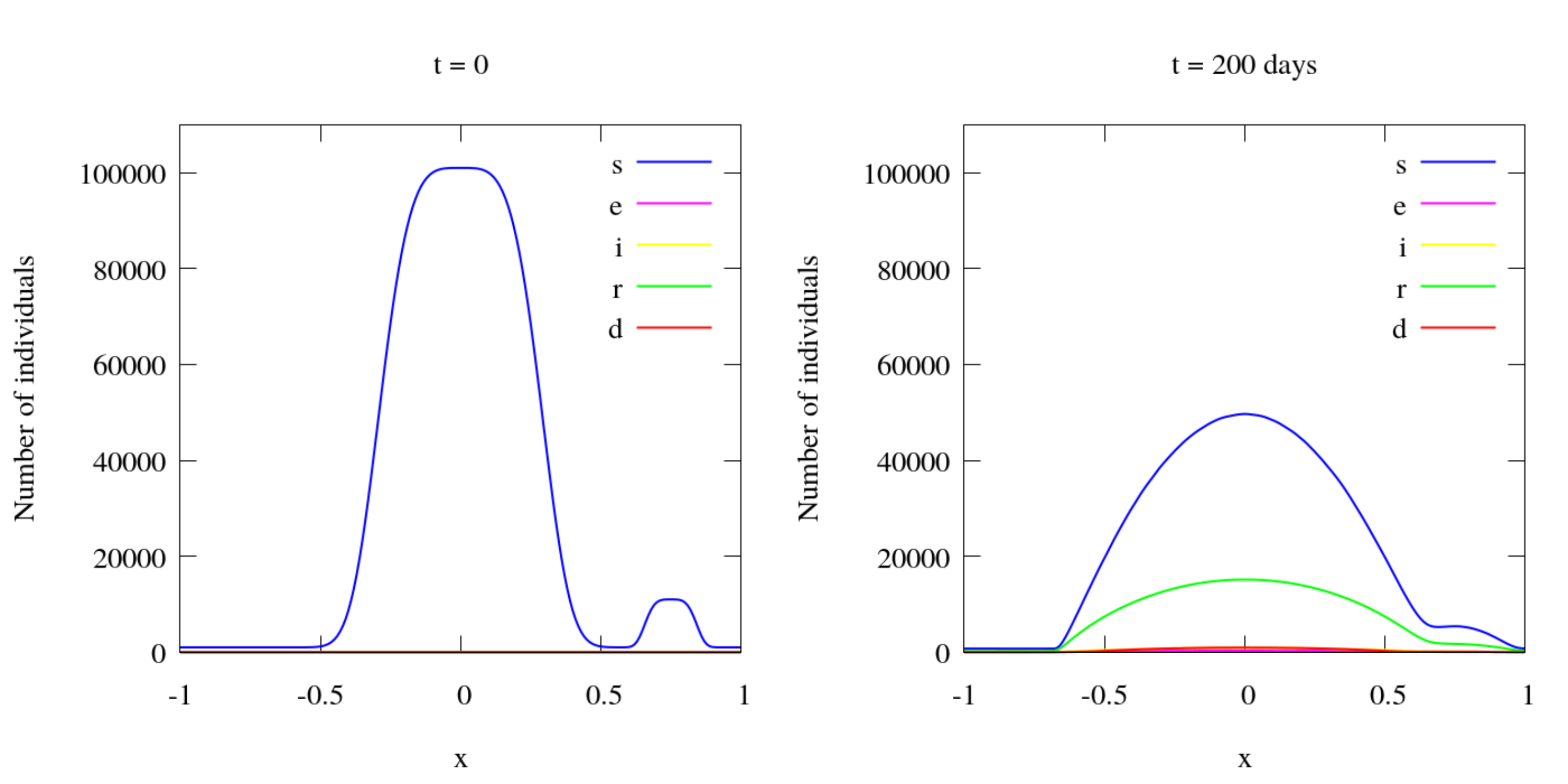}
    \caption{COVID Test 5: Populations over a horizontal line crossing the middle of the domain.}
    \label{fig:xx-source}
\end{figure}

\begin{figure}[htpb]
    \centering
    \includegraphics[width=0.9\textwidth]{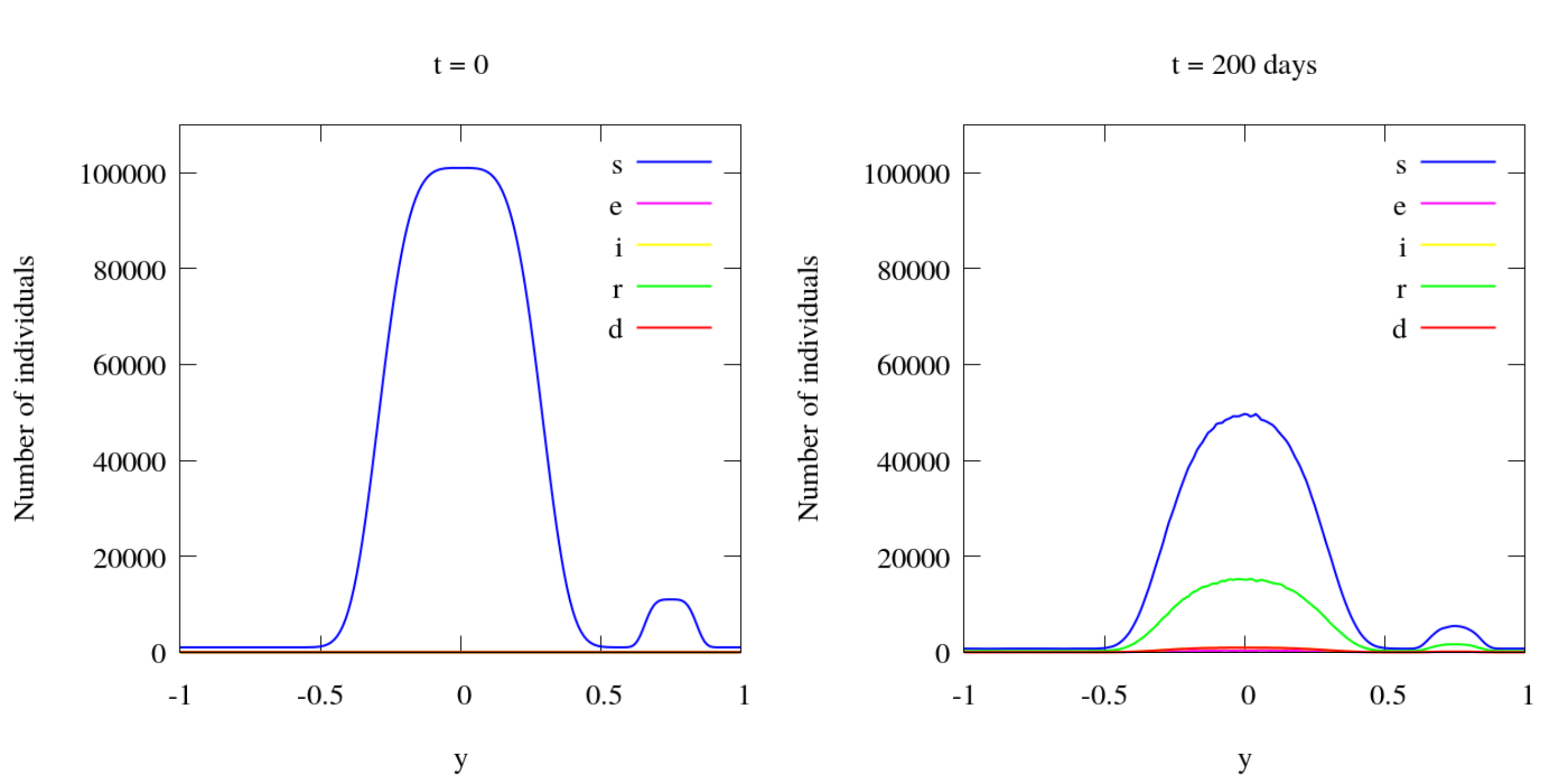}
    \caption{COVID Test 5: Populations over a vertical line crossing the middle of the domain.}
    \label{fig:yy-source}
\end{figure}

Figure \ref{fig:source-time} shows the populations at different time-steps. We see oscillations in the number of individuals of the populations coming from the random source dynamic. These oscillations are smoothed in the $x$ direction because of the diffusion. We can see this better in  Figures \ref{fig:xx-source} and \ref{fig:yy-source} that shows the results over a centralized horizontal and vertical line crossing the domain, respectively. The vertical plot shows unsmoothed oscillations coming from the random source in the $y$ direction. In this example, it is possible to better seeing the effects of anisotropic diffusion. Note that in the horizontal plot, the populations spread over the $x$ direction, while in the vertical plot, the populations change the compartments but stay in the same coordinates.  

\begin{figure}[htpb]
    \centering
    \includegraphics[width=0.5\textwidth]{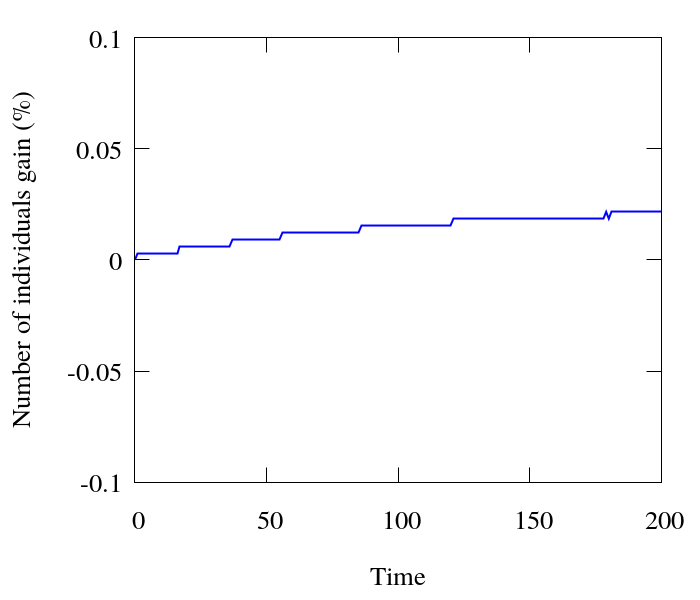}
    \caption{COVID Test 5: Time history of the total number of individuals.}
    \label{fig:mass_source}
\end{figure}

In Figure \ref{fig:mass_source}, we plot the time history of the total number of individuals. There is a negligible increase in the total number of individuals (less than 0.1\%). 

\section{Conclusions}

We developed an extended continuum SEIRD model to represent the dynamics of the COVID-19 virus spread based on the framework proposed in \cite{viguerie2020simulating}. We validate our code by comparing our results with other simulations. We introduce new test cases to highlight new modeling capabilities. Among the new features added to the base model, there is the addition of a source term, which represents exposed people who return from travel, by changing individuals from the susceptible compartment to the exposed compartment. We also add the possibility of anisotropic non-homogeneous diffusion. Our code is implemented through the \texttt{libMesh} library and supports adaptive mesh refinement and coarsening. Therefore, it can represent several spatial scales, adapting the resolution to the disease dynamics.  

Data is essential to define the epidemic spreading parameters, as diffusion and infection rate. We have to study how it would be the best way to represent people who return from travels, addressing questions like defining the probability of a random source appearing in the system, in which area, the population density, among others. Diffusion-reaction models, as the present one, are richer than standard compartmental models. However, they are slower, which hampers their widespread utilization in what-if scenarios, parametric studies, and time-critical situations. Therefore, the development of low-dimensional computational models will leverage the ability of continuous models to perform in real-time scenarios. Projection-based or data-driven model order reduction \cite{ROMbook,DMDbook} aims to lower the computational complexity of a given computational model by reducing its dimensionality (or order), can provide this leverage. They can work in conjunction with emerging machine learning methods such as physics informed neural networks \cite{raissi2019physics}. We can foresee a tremendous impact in the mathematical epidemiology field of all these new methods and techniques, enlarging the predictive capabilities and computational efficiency of diffusion-reaction epidemiological models.

\section*{Acknowledgements}
This research was financed in part by the Coordena\c{c}\~ao de Aperfei\c{c}oamento de Pessoal de N\'ivel Superior - Brasil (CAPES) - Finance Code 001 and CAPES TecnoDigital Project 223038.014313/2020-19.  This research has also received funding from CNPq and FAPERJ. We are indebted to Prof. Americo Cunha Jr., Prof. Regina Almeida and Prof. Sandra Malta for fruitful discussions and invaluable help in the understanding of epidemiological models.


\begin{appendix}

\section{Implementation of the generic spatio-temporal SEIRD model}

We implement the generic SEIRD model similar to the EPIDEMIC software. We have used the BDF2 time discretization method, Newton's method for the nonlinear terms, and we simplify the number of the living population by considering the previous time-step solution. For all test cases the nonlinear tolerance for Newton's method is set to $10^{-8}$ and the linear solver tolerance is set to $10^{-10}$. The linear solver is GMRES with ILU(0) preconditioner. 

In \texttt{libMesh}, we calculate directly the new solution ($\mathbf{u}_{n+1})$ instead of the variation ($\delta \mathbf{u})$. Then, on the left-hand side, we gather the terms containing an unknown, whereas all the other terms are taken to the right-hand-side. The superscript $k$ is from the previous Newton iteration.
The terms in black are from the mass matrix, in blue are the nonlinear terms, in red the diffusive terms, and in green the remaining terms from the stiffness matrix. The finite element shape functions are represented by $N_a$, $a= 1, \cdots , n_{nnos}$, where $n_{nnos}$ is the number of nodes in the finite element mesh.

Susceptible (Equation \ref{epidemic_s}): 
\begin{equation}
     K_{ss} = \int_{\Omega_e} 1.5N_a N_b  d\Omega  \textcolor{blue}{+ \Delta t \int_{\Omega_e}  N_a \beta N_b \frac{i_k}{n_{k}}} \textcolor{red}{+ \Delta t \int_{\Omega_e} \nabla N_a n_{k} \nu_s \nabla N_b d\Omega}
\end{equation}

\begin{equation}
     K_{si} = \textcolor{blue}{\Delta t \int_{\Omega_e}  N_a \beta N_b \frac{s_k}{n_{k}} d\Omega}
\end{equation}

\begin{equation}
     F_{s} = \int_{\Omega_e}  N_a (2s_{n}-0.5s_{n-1}) d\Omega  \textcolor{blue}{+ \Delta t \int_{\Omega_e}  N_a \beta \frac{s_k i_k}{n_{k}} d\Omega}
\end{equation}

Exposed (Equation \ref{epidemic_e}):

\begin{equation}
     K_{ee} = \int_{\Omega_e}1.5 N_a N_b  d\Omega  \textcolor{green}{+\Delta t \int_{\Omega_e} \alpha N_a N_b  d\Omega}  \textcolor{red}{+\Delta t \int_{\Omega_e} \nabla N_a n_{old}\nu_e \nabla N_b d\Omega}
\end{equation}

\begin{equation}
     K_{ei} = \textcolor{blue}{- \Delta t \int_{\Omega_e}  N_a \beta N_b \frac{s_k}{n_{k}} d\Omega}
\end{equation}

\begin{equation}
     K_{es} = \textcolor{blue}{ - \Delta t \int_{\Omega_e}  N_a \beta N_b \frac{i_k}{n_{k}} d\Omega}
\end{equation}

\begin{equation}
     F_{e} = \int_{\Omega_e}  N_a (2e_{n}-0.5e_{n-1}) d\Omega \textcolor{blue}{- \Delta t \int_{\Omega_e}  N_a \beta \frac{s_k i_k}{n_{k}} d\Omega}
\end{equation}

Infected (Equation \ref{epidemic_i}):

\begin{equation}
     K_{ii} = \int_{\Omega_e} 1.5N_a N_b  d\Omega \textcolor{green}{+ \Delta t \int_{\Omega_e} (\gamma + \delta) N_a N_b  d\Omega} \textcolor{red}{ +  \Delta t \int_{\Omega_e} \nabla N_a n_{k} \nu_i \nabla N_b d\Omega}
\end{equation}

\begin{equation}
     K_{ie} = \textcolor{green}{- \Delta t \int_{\Omega_e} \alpha N_a N_b  d\Omega}
\end{equation}

\begin{equation}
     F_{i} = \int_{\Omega_e}  N_a (2i_{n}-0.5i_{n-1}) d\Omega 
\end{equation}

Recovered (Equation \ref{epidemic_r}):

\begin{equation}
     K_{rr} = \int_{\Omega_e} 1.5N_a N_b  d\Omega \textcolor{red}{+  \Delta t \int_{\Omega_e} \nabla N_a n_{k} \nu_r \nabla N_b d\Omega}
\end{equation}

\begin{equation}
     K_{ri} = \textcolor{green}{- \Delta t\int_{\Omega_e} \gamma N_a N_b  d\Omega}
\end{equation}

\begin{equation}
     F_{r} = \int_{\Omega_e}  N_a (2r_{n}-0.5r_{n-1}) d\Omega 
\end{equation}

Diseased (Equation \ref{epidemic_d}):

\begin{equation}
     K_{dd} = \int_{\Omega_e} 1.5N_a N_b  d\Omega
\end{equation}

\begin{equation}
     K_{di} = \textcolor{green}{- \Delta t \int_{\Omega_e} \delta N_a N_b  d\Omega}
\end{equation}

\begin{equation}
     F_{d} = \int_{\Omega_e}  N_a (2d_{n}-0.5d_{n-1}) d\Omega 
\end{equation}

\section{Implementation of the spatio-temporal model of COVID-19 infection spread}

We present the matrix contributions of the system of equations that represents the COVID19 dynamics \cite{viguerie2020simulating,viguerie2020diffusion}. We use the BDF2 time discretization method, Newton's method for the nonlinear terms, and we simplify the number of the living population by considering the previous linear solution. For all test cases the nonlinear tolerance for Newton's method is set to $10^{-8}$ and the linear solver tolerance is set to $10^{-10}$. The linear solver is GMRES with ILU(0) preconditioner.

In \texttt{libMesh}, we calculate directly the new solution ($\mathbf{u}_{n+1})$ instead of the variation ($\delta \mathbf{u})$. Then, on the left-hand side, we gather the terms containing an unknown, whereas all the other terms are taken to the right-hand-side. The superscript $k$ is from the previous Newton iteration.
The terms in black are from the mass matrix, in blue are the nonlinear terms, in red the diffusive terms, in green the remaining terms from the stiffness matrix and in yellow the source terms.

Susceptible (Equation \ref{covid_s}): 
\begin{equation}
\begin{split}
     K_{ss} = \int_{\Omega_e} 1.5N_a N_b  d\Omega  \textcolor{blue}{+ \Delta t \int_{\Omega_e}  N_a \beta_i \left(1-\frac{A}{n_k}\right) N_b i_k + \Delta t \int_{\Omega_e}  N_a \beta_e \left(1-\frac{A}{n_k}\right) N_be_k d\Omega}\\
     \textcolor{red}{+ \Delta t \int_{\Omega_e} \nabla N_a n_k \nu_s \nabla N_b d\Omega}
     \end{split}
\end{equation}

\begin{equation}
     K_{si} = \textcolor{blue}{\Delta t \int_{\Omega_e}  N_a \beta_i \left(1-\frac{A}{n_k}\right) N_b s_k d\Omega}
\end{equation}

\begin{equation}
     K_{se} = \textcolor{blue}{\Delta t \int_{\Omega_e}  N_a \beta_e \left(1-\frac{A}{n_k}\right) N_b s_k d\Omega}
\end{equation}

\begin{equation}
\begin{split}
     F_{s} = \int_{\Omega_e}  N_a (2s_{n}-0.5s_{n-1}) d\Omega  \textcolor{blue}{+ \Delta t \int_{\Omega_e}  N_a \beta_i \left(1-\frac{A}{n_k}\right) s_k i_k d\Omega 
     + \Delta t \int_{\Omega_e}  N_a \beta_e \left(1-\frac{A}{n_k}\right) s_k e_k d\Omega} \textcolor{yellow}{+f}
\end{split}
\end{equation}

Exposed (Equation \ref{covid_e}):

\begin{equation}
\begin{split}
     K_{ee} = \int_{\Omega_e}1.5 N_a N_b  d\Omega  \textcolor{green}{+\Delta t \int_{\Omega_e} (\alpha+\gamma_e) N_a N_b  d\Omega}  \textcolor{red}{+\Delta t \int_{\Omega_e} \nabla N_a n_k\nu_e \nabla N_b d\Omega} \\ \textcolor{blue}{- \Delta t \int_{\Omega_e}  N_a \beta_e \left(1-\frac{A}{n_k}\right) N_b s_k d\Omega}
     \end{split}
\end{equation}

\begin{equation}
     K_{ei} = \textcolor{blue}{- \Delta t \int_{\Omega_e}  N_a \beta_i  \left(1-\frac{A}{n_k}\right) N_b s_k d\Omega}
\end{equation}

\begin{equation}
     K_{es} = \textcolor{blue}{ - \Delta t \int_{\Omega_e}  N_a \beta_i \left(1-\frac{A}{n_k}\right) N_b i_k d\Omega- \Delta t \int_{\Omega_e}  N_a \beta_e \left(1-\frac{A}{n_k}\right) N_b e_k d\Omega}
\end{equation}

\begin{equation}
\begin{split}
     F_{e} = \int_{\Omega_e}  N_a (2e_{n}-0.5e_{n-1}) d\Omega \textcolor{blue}{- \Delta t \int_{\Omega_e}  N_a \beta_i \left(1-\frac{A}{n_k}\right) s_k i_k d\Omega} \\ \textcolor{blue}{- \Delta t \int_{\Omega_e}  N_a \beta_e \left(1-\frac{A}{n_k}\right) s_k e_k d\Omega} \textcolor{yellow}{-f}
     \end{split}
\end{equation}

Infected (Equation \ref{covid_i}):

\begin{equation}
     K_{ii} = \int_{\Omega_e} 1.5N_a N_b  d\Omega \textcolor{green}{+ \Delta t \int_{\Omega_e} (\gamma_i + \delta) N_a N_b  d\Omega} \textcolor{red}{ +  \Delta t \int_{\Omega_e} \nabla N_a n_k \nu_i \nabla N_b d\Omega}
\end{equation}

\begin{equation}
     K_{ie} = \textcolor{green}{- \Delta t \int_{\Omega_e} \alpha N_a N_b  d\Omega}
\end{equation}

\begin{equation}
     F_{i} = \int_{\Omega_e}  N_a (2i_{n}-0.5i_{n-1}) d\Omega 
\end{equation}

Recovered (Equation \ref{covid_r}):

\begin{equation}
     K_{rr} = \int_{\Omega_e} 1.5N_a N_b  d\Omega \textcolor{red}{+  \Delta t \int_{\Omega_e} \nabla N_a n_k \nu_r \nabla N_b d\Omega}
\end{equation}

\begin{equation}
     K_{ri} = \textcolor{green}{- \Delta t\int_{\Omega_e} \gamma_i N_a N_b  d\Omega}
\end{equation}

\begin{equation}
     K_{re} = \textcolor{green}{- \Delta t\int_{\Omega_e} \gamma_e N_a N_b  d\Omega}
\end{equation}

\begin{equation}
     F_{r} = \int_{\Omega_e}  N_a (2r_{n}-0.5r_{n-1}) d\Omega 
\end{equation}

Diseased (Equation \ref{covid_d}):

\begin{equation}
     K_{dd} = \int_{\Omega_e} 1.5N_a N_b  d\Omega
\end{equation}

\begin{equation}
     K_{di} = \textcolor{green}{- \Delta t \int_{\Omega_e} \delta N_a N_b  d\Omega}
\end{equation}

\begin{equation}
     F_{d} = \int_{\Omega_e}  N_a (2d_{n}-0.5d_{n-1}) d\Omega 
\end{equation}

\end{appendix}


\bibliographystyle{unsrt}  
\bibliography{references}  

\end{document}